\documentclass[a4paper,aps,pre,showpacs,floatfix,twocolumn,%
superscriptaddress,nobibnotes,showkeys]{revtex4-1}
\usepackage{graphics}
\usepackage{graphicx}
\usepackage{psfrag}
\usepackage{latexsym}
\usepackage{textcomp}
\usepackage{amssymb}
\usepackage{amsmath}
\usepackage{bm}
\usepackage[Euler]{upgreek}
\usepackage{longtable}
\usepackage{subfigure}
\usepackage{epsfig}
\usepackage{dcolumn}

\DeclareMathAlphabet{\mathsf}{OT1}{phv}{b}{n}

\newcommand{\Vector}[1]{\ensuremath{\mathbf{#1}}}
\newcommand{\Tensor}[1]{\ensuremath{\mathsf{#1}}}

\newcommand{\bcal}[1]{\bm{\mathcal{#1}}}

\newcommand{\avg}[1]{\left< #1 \right>}

\newcommand{\erfc}{\text{erfc}}

\newcommand{\crossVorg}{\ensuremath{%
         \setbox0=\hbox{$V$}
        V \kern-\wd0{\raise.3ex\hbox{$\relbar$}}}}

\newcommand{\crossVxx}[2]{%
	{\setbox0=\hbox{$#1#2V$}
         \setbox1=\hbox{$#1#2$}
         \setbox2=\hbox{$#1V$}
         \dimen1=\wd0
	 \advance\dimen1-\wd1
         \raise.2\ht0\hbox{$#1#2$}\kern-.4\wd0}}

\newcommand{\ie}{i.e.}

\usepackage{pifont}

\begin{document}
\title{Optimisation of a Brownian dynamics algorithm for
semidilute polymer solutions}
\date{\today}
\author{Aashish Jain}
\affiliation{Department of Chemical Engineering, Monash University,
Melbourne, VIC 3800, Australia}
\author{P. Sunthar}
\affiliation{Department of Chemical Engineering, Indian Institute of 
Technology Bombay, Powai, Mumbai - 400076, India}
\author{B. D\"{u}nweg}
\affiliation{Max Planck Institute for Polymer Research,
Ackermannweg 10, 55128 Mainz, Germany}
\affiliation{Department of Chemical Engineering, Monash University,
Melbourne, VIC 3800, Australia}
\author{J. Ravi Prakash}
\email{Email: ravi.jagadeeshan@monash.edu \\
URL: http://users.monash.edu.au/~rprakash/}
\affiliation{Department of Chemical Engineering, Monash University,
Melbourne, VIC 3800, Australia}

\date{\today}

\begin{abstract}
  Simulating the static and dynamic properties of semidilute polymer
  solutions with Brownian dynamics (BD) requires the computation of a
  large system of polymer chains coupled to one another through
  excluded-volume and hydrodynamic interactions.  In the presence of
  periodic boundary conditions, long-ranged hydrodynamic interactions
  are frequently summed with the Ewald summation technique. By
  performing detailed simulations that shed light on the influence of
  several tuning parameters involved both in the Ewald summation
  method, and in the efficient treatment of Brownian forces, we
  develop a BD algorithm in which the computational cost scales as
  $O(N^{1.8})$, where $N$ is the number of monomers in the simulation
  box. We show that Beenakker's original implementation of the Ewald
  sum, which is only valid for systems without bead overlap, can be
  modified so that $\theta$-solutions can be simulated by switching
  off excluded-volume interactions. A comparison of the predictions of
  the radius of gyration, the end-to-end vector, and the
  self-diffusion coefficient by BD, at a range of concentrations, with
  the hybrid Lattice Boltzmann/Molecular Dynamics (LB/MD) method shows
  excellent agreement between the two methods. In contrast to the
  situation for dilute solutions, the LB/MD method is shown to be
  significantly more computationally efficient than the current
  implementation of BD for simulating semidilute solutions. We argue
  however that further optimisations should be possible.
\end{abstract}

\maketitle

\section{\label{sec:intro}Introduction}

Understanding the behaviour of polymer solutions in the semidilute
regime of concentration is important both from the point of view of
applications, and from the point of view of advancing fundamental
knowledge in polymer science. Until recently, insight into semidilute
polymer solution behaviour was largely obtained through approximate
analytical and scaling theories~\cite{de-gennes-macro1976-1,%
  de-gennes-paper,de-gennes,muthukumar1982polymer,muthukumar1982JCP,%
  muthukumar1983macro,muthukumar1984macro,rubinstein}. However,
significant progress in the development of mesoscopic simulation
techniques~\cite{BD2001,stoltz,gompper2010}, which allow the
exploitation of underlying theories without the need for
approximations, has made it possible for the first time to obtain
detailed predictions of equilibrium and nonequilibrium properties that
can be compared with experimental observations. The successful
implementation of mesoscopic simulations has been made possible
through the use of algorithms that enable an accurate depiction of the
semidilute regime. Essentially, this requires the ability to describe
long polymers that overlap with each other, while maintaining a low
segment density. Further, the segments must be capable of interacting
with each other through solvent-mediated hydrodynamic
interactions~\cite{kirkwood1948,edwards-freedJCP1974,de-gennes-paper,%
  bixon1976,BD2001}. Three different mesoscopic simulation methods,
all of which use coarse-grained bead-spring chain models for polymer
molecules, have been developed recently that achieve these
objectives. Two of these techniques, namely, the hybrid Lattice
Boltzmann/Molecular Dynamics (LB/MD) method~\cite{BD1999,BDTL2009} and
the hybrid Multiparticle Collision Dynamics/Molecular Dynamics (MPCD)
method~\cite{kapral1999JCP,kapral_bookchapter,gompper2009} treat the
solvent explicitly. As a consequence, hydrodynamic interactions
between polymer segments arise naturally through the exchange of
momentum between the beads on a chain and solvent molecules. In the
third approach~\cite{stoltz}, which is Brownian dynamics (BD)
simulations~\cite{ottinger}, the solvent degrees of freedom are
removed completely, but their effect is taken into account through
long-range dynamic correlations in the stochastic displacements of the
beads. The very nature of semidilute polymer solutions, particularly
the need to use periodic boundary conditions to describe homogeneous
polymer solutions in unbounded domains, necessitates the simulation of
a large number of particles. As a result, the computational efficiency
of a simulation technique becomes an important consideration.

To our knowledge, there has been no systematic investigation to
compare the performance of the different techniques in terms of their
computational efficiency in the semidilute regime. Recently, however,
a quantitative comparison of the predictions of the explicit solvent
LB/MD method with the predictions of the implicit solvent BD method
for the dynamics of a single chain in a solvent, \ie\ in the dilute
regime, has been carried out with a view to compare their
computational efficiencies~\cite{tri,Butler2009PRE}. It was shown by
\citet{tri} that in order to observe the system for the same time span
in physical units, significantly less CPU time is required with BD in
comparison to LB/MD, for bead-spring chains with $N_{b} \alt 10^{6}$,
where $N_{b}$ is the number of beads per chain. The situation,
however, is expected to be quite different in the semidilute
regime. For the LB/MD method, the CPU cost scales linearly with the
number of particles, which implies that the CPU cost grows as $L^{3}$
since the solvent particles (the calculation of whose dynamics
dominates the CPU cost), are distributed on lattice grid points in a
simulation box of size $L$. In order to prevent a chain from wrapping
over itself due to spatial restriction and hence altering its static
conformation, it is necessary to ensure that $L \ge 2
\sqrt{\avg{R_{e}^{2}}}$, where $\avg{R_{e}^{2}}$ is the mean-square
end-to-end distance of the chain. In the dilute case, this leads to
the CPU time scaling as $N_{b}^{3\nu}$ for the LB/MD method, where
$\nu$ is the Flory exponent. Using a simple scaling argument based on
the {blob} picture of semidilute solutions~\cite{de-gennes},
\citet{tri} suggest that the CPU effort is even somewhat decreased in
semidilute solutions due to the shrinkage of the chains resulting from
the screening of excluded-volume
interactions~\cite{de-gennes,doi-edwards-book} --- note that this
argument is based upon considering the smallest possible system size.

In the case of BD, even though the number of degrees of freedom is
significantly reduced by eliminating the solvent, implementation of
pairwise hydrodynamic interactions between segments proves to be
extremely computationally expensive. For dilute polymer solutions, the
computational cost of evaluating intramolecular hydrodynamic
interactions arises from the need to carry out a decomposition of the
diffusion tensor that appears in the stochastic equation of motion. A
straightforward Cholesky decomposition leads to an algorithm that
scales like $O(N_{b}^{3})$. Many current implementations of single
chain BD simulations, however, mitigate this large CPU cost by using
Fixman's polynomial approximation to this decomposition, which leads
to $O(N_{b}^{2.25})$
scaling~\cite{fixman,jendrejack,kroger,prabhakar}. In the case of
semidilute polymer solutions, both intramolecular and intermolecular
hydrodynamic interactions must be taken into account. The use of
periodic boundary conditions to imitate bulk systems necessitates the
evaluation of hydrodynamic interactions not only between one
particular chosen segment and all other segments within the primary
simulation box, but also with all the other segments in all the
periodic images of the box. Because of the long-range nature of
hydrodynamic interactions, which decay only reciprocally with
distance, this sum converges very slowly and only
conditionally~\cite{hasimoto,beenakker}. Inspired by its earlier
success in summing electrostatic interaction (which are also
long-ranged in nature), the problem of slow convergence has been
resolved through the use of the Ewald summation
technique~\cite{ewald,allen,luty,toukmaji,holm} --- both in the
context of BD simulations of colloidal suspensions where hydrodynamic
interactions between particles with a finite radius must be taken into
account~\cite{snook1987,bplb,maret1999,sierou,banchio2003}, and in the
context of BD simulations of semidilute polymer solutions where the
polymer segments are assumed to be point particles~\cite{stoltz}.

Rapid convergence is achieved in the Ewald method by splitting the
original expression into two sums, one of them in real space and the
other in reciprocal space, both of which converge exponentially. A
straightforward implementation of the Ewald sum, however, is
computationally demanding, scaling like $O(N^{2})$, where $N = N_{c}
\times N_{b}$, is the total number of beads in the primary simulation
box with $N_{c}$ polymer chains. Interestingly, by a suitable choice
of a parameter $\alpha$ in the Ewald sum that tunes the relative
weights of the real space and reciprocal space contributions
(consequently splitting the load of calculating the total sum between
the real space and reciprocal space sums), it is possible to make the
computational cost of calculating either the real space or the
reciprocal space sum scale like $O(N^{2})$, while the remaining sum
scales as $O(N)$. In their recent simulation of semidilute polymer
solutions, \citet{stoltz} have implemented a BD algorithm that leads
to the real space sum scaling like $O(N^{2})$. In the case of
colloidal suspensions, accelerated BD algorithms have been developed
by Brady and co-workers with the Ewald sum scaling like $O(N \log
N)$~\cite{sierou,banchio2003}. The essential idea is to retain an
$O(N)$ scaling for the real space sum, while reducing the complexity
of the reciprocal part of the Ewald sum to $O(N \log N)$ with the help
of Fast Fourier Transformation. For confined systems which are
non-periodic, and in which methods based on Fourier transforms are not
applicable, the Wisconsin group have recently successfully introduced
a BD simulation technique they term the ``general geometry Ewald-like
method'', which achieves $O(N \log N)$ scaling~\cite{ggem-graham2012}.
Analogous to the Ewald method, the technique is based on splitting the
solution to the Stokes equation into singular short-ranged parts and
smooth long-ranged parts. Thus, even though a detailed quantitative
comparison of all the currently available mesoscopic simulation
techniques is yet to be carried out, they all appear to scale, in
their most efficient versions, roughly linearly with system size.

In the context of electrostatic interactions, two different classes of
schemes have been proposed for the optimisation of the Ewald
sum~\cite{allen,holm,perram,fincham}. One of these classes (on which
the accelerated BD schemes are modelled), achieves $O(N \log N)$
scaling by assigning particles to a mesh and then using Fast Fourier
Transform techniques to evaluate the reciprocal-space part of the
Ewald sum on this mesh~\cite{luty,holm}. The other
class~\cite{perram,fincham} achieves $O(N^{1.5})$ scaling by balancing
the computational cost of evaluating the real space and reciprocal
space sums, \ie\ by an optimal choice of the aforementioned parameter
$\alpha$. To our knowledge, the latter approach has so far not been
trialled for summing hydrodynamic interactions.

In the context of hydrodynamic interactions, it is also worth noting
that the BD simulation of semidilute polymer solutions carried out by
\citet{stoltz} differs from BD simulations of colloidal
suspensions~\cite{bplb,banchio2003} in the procedure adopted for the
calculation of \emph{far-field} hydrodynamic interactions, even though
both are based on the Ewald summation technique. While the latter are
based on Hasimoto's solution of the Stokes equation for flow past a
periodic array of point forces~\cite{hasimoto}, \ie\ on the Ewald sum
of the Oseen-Burgers tensor, the former is based on Beenakker's
solution~\cite{beenakker}, which is the Ewald sum of the
Rotne-Prager-Yamakawa (RPY) tensor. The RPY
tensor~\cite{rpy,yamakawa70,ottinger} is a generalisation of the
Oseen-Burgers tensor in two aspects: Firstly, it approximately takes
into account the finite particle radius by representing the far-field
hydrodynamic flow field up to quadrupolar order of its multipole
expansion~\cite{MazurVanSaarloos} (Oseen-Burgers is just a monopole
field), and, secondly, it regularises the singularity that occurs for
small inter-bead distances. Such a regularisation is necessary for BD
simulations that allow for configurations with overlapping beads,
since otherwise the diffusion tensor would not always be
positive-definite, implying a violation of the second law of
thermodynamics. The problem can be avoided by introducing sufficiently
strong excluded-volume interactions, which suppress the occurrence of
such configurations; this was the approach taken by \citet{stoltz}. By
this procedure, they also avoided the problem that Beenakker's
formulae~\cite{beenakker} are applicable only to the far-field branch
of the RPY tensor and not to the regularised near-field
branch. However, one can anticipate that bead overlap will occur in
simulations of semidilute $\theta$-solutions, since $\theta$-solutions
are commonly simulated by switching off excluded-volume
interactions~\cite{prabhakar,suntharmacro}. We therefore develop in
the present paper a method that is able to deal with overlaps, by
implementing the Ewald sum of \emph{both branches} of the RPY
tensor. It should be noted that the near-field RPY formula is not the
only possible regularisation that has been discussed in the
literature; an alternative was suggested in Ref.~\onlinecite{zylka89}.
While the details of the regularisation are unlikely to significantly
affect the dynamic properties of semidilute polymer solutions, we have
focused on the RPY formula, since, according to our knowledge, it is
the only known regularisation that provenly provides
positive-definiteness for all chain lengths and
configurations~\cite{ottinger}. For a physical motivation of the RPY
regularisation, see Ref.~\cite{rpy}. In the context of BD simulations
of colloidal suspensions, the problem of positive-definiteness does
not arise due to excluded-volume interactions, while \emph{near-field}
hydrodynamic interactions are taken into account through short-range
lubrication forces~\cite{bplb}.

In this paper, four aspects of the implementation and optimisation of
BD simulations for semidilute polymer solutions are considered: (i)
The development of an algorithm that scales like $O(N^{1.5})$ for
calculating pair-wise hydrodynamic interactions in periodic systems;
(ii) The derivation of a modified version of Beenakker's periodic RPY
tensor that is applicable to simulations of solutions under $\theta$
conditions; (iii) The optimal implementation of Fixman's polynomial
approximation to the decomposition of the diffusion tensor, within the
context of the current BD simulation algorithm and (iv) Optimisation
of the overall algorithm for a single Euler time step. Finally, the
resulting optimised BD algorithm has been used to calculate a variety
of equilibrium properties of semidilute polymer solutions, across a
range of concentrations, and compared quantitatively with the results
of the LB/MD algorithm, along with a comparison of the CPU time
scaling for both these approaches.

The plan of the paper is as follows. In Sec.~\ref{sec:model}, the
governing equations, along with the implementation of the Ewald sum
and its modification to handle overlapping beads, are
discussed. Sections~\ref{sec:opt-DF}, \ref{sec:chebyshev},
and~\ref{sec:euler-opt} consider the optimisation of (i) the Ewald sum
for hydrodynamic interactions, (ii) the Chebychev polynomial
approximation for the decomposition of the diffusion tensor, and (iii)
the execution of a single Euler time step, respectively. The optimised
BD algorithm is validated by a variety of different means, under both
$\theta$ and good solvent conditions, in
Sec.~\ref{sec:prop-validation}, and in
Sec.~\ref{sec:CPUTIME-COMPARISON}, its computational cost is
compared with that of the LB/MD method at a concentration that lies in
the semidilute regime. Finally, the principal conclusions are
summarised in Sec.~\ref{sec:summary}.

\section{\label{sec:model} Model and simulation method}
\subsection{\label{sec:brownian} Governing equation}

A linear bead-spring chain model is used to represent polymers at the
mesoscopic level, with each polymer chain coarse-grained into a
sequence of $N_b$ beads, which act as centres of hydrodynamic
resistance, connected by $N_b - 1$ massless springs that represent the
entropic force between adjacent beads. A semidilute polymer solution
is modelled as an ensemble of such bead-spring chains, immersed in an
incompressible Newtonian solvent. A total of $N_c$ chains are
initially enclosed in a cubic and periodic cell of edge length $L$,
giving a total of $N = N_b \times N_c$ beads per cell at a bulk
monomer concentration of $c = N/V$, where $V = L^3$ is the volume of
the simulation cell. Using the length scale $l_H = \sqrt{k_B T/H}$ and
time scale ${\lambda}_H = \zeta/4H$, where $k_B$ is the Boltzmann's
constant, $T$ is the temperature, $H$ is the spring constant and
$\zeta$ is the hydrodynamic friction coefficient associated with a
bead, the Euler integration algorithm for the non-dimensional Ito
stochastic differential equation governing the position vector
${\Vector{r}}_{\nu}(t)$ of bead $\nu$ at time $t$, is~\cite{stoltz}
\begin{eqnarray}
\label{eq:sde}
\nonumber
{\Vector{r}}_{\nu}(t+\Delta t)  
& = &
{\bm{\Vector{r}}}_{\nu}(t) + \left[\bm{\Tensor{\kappa}} 
\cdot \bm{\Vector{r}}_{\nu}(t) \right] 
+ \frac{\Delta t}{4} \, \sum_{\mu = 1}^{N} 
\left [{\bm{\Tensor{D}}}_{\nu \mu}(t) \cdot {\bm{\Vector{F}}}_{\mu}(t)\right] 
\\
& +  &
\frac{1}{\sqrt{2}} \, \sum_{\mu = 1}^{N} 
\left [{\bm{\Tensor{B}}}_{\nu \mu}(t) 
\cdot {\Vector{\Delta \bm{W}}}_{\mu}(t) \right]
\end{eqnarray}
Here, the $3 \times 3$ tensor $\bm{\Tensor{\kappa}}$ is equal to
${(\bm{\nabla v})}^T$, with $\bm{v}$ being the unperturbed solvent
velocity. The dimensionless diffusion tensor $\bm{\Tensor{D}}_{\nu
  \mu}$ is a $3 \times 3$ matrix for a fixed pair of beads $\mu$ and
$\nu$. It is related to the hydrodynamic interaction tensor, as
discussed further subsequently. ${\bm{\Vector{F}}}_{\mu}$ incorporates
all the non-hydrodynamic forces on bead $\mu$ due to all the other
beads. The components of the Gaussian noise $\Vector{\Delta
  \bm{W}}_{\mu}$ are obtained from a real-valued Gaussian distribution
with zero mean and variance $\Delta t$. The quantity
$\bm{\Tensor{B}}_{\nu \mu}$ is a non-dimensional tensor whose presence
leads to multiplicative noise~\cite{ottinger}. Its evaluation requires
the decomposition of the diffusion tensor. Defining the matrices
$\bcal{D}$ and $\bcal{B}$ as block matrices consisting of $N \times N$
blocks each having dimensions of $3 \times 3$, with the $(\nu,
\mu)$-th block of $\bcal{D}$ containing the components of the
diffusion tensor $\bm{\Tensor{D}}_{\nu \mu}$, and the corresponding
block of $\bcal{B}$ being equal to $\bm{\Tensor{B}}_{\nu \mu}$, the
decomposition rule for obtaining $\bcal{B}$ can be expressed as
\begin{gather}
\bcal{B} \cdot {\bcal{B}}^\textsc{t} = \bcal{D} \label{decomp}
\end{gather}
The non-hydrodynamic forces in the model are comprised of the spring
forces ${\bm{\Vector{F}}}_{\mu}^{\text{spr}}$ and excluded-volume
interaction forces ${\bm{\Vector{F}}}_{\mu}^{\text{exv}}$, \ie,
${\bm{\Vector{F}}}_{\mu} = {\bm{\Vector{F}}}_{\mu}^{\text{spr}} +
{\bm{\Vector{F}}}_{\mu}^{\text{exv}}$. A linear Hookean spring
potential is used here for modelling the spring forces when
considering the optimisation of the Ewald sum, while a finitely
extensible nonlinear elastic (FENE) potential has been used while
comparing results with the Lattice Boltzmann method. The entropic
spring force on bead $\mu$ due to adjacent beads can be expressed as
${\bm{\Vector{F}}}_{\mu}^{\text{spr}} =
{\bm{\Vector{F}}}^c({\bm{\Vector{Q}}}_{\mu}) -
{\bm{\Vector{F}}}^c({\bm{\Vector{Q}}}_{\mu - 1})$ where
${\bm{\Vector{F}}}^c({\bm{\Vector{Q}}}_{\mu - 1})$ is the force
between the beads $\mu -1$ and $\mu$, acting in the direction of the
connector vector between the two beads ${\bm{\Vector{Q}}}_{\mu - 1} =
{\bm{\Vector{r}}}_{\mu} - {\bm{\Vector{r}}}_{\mu - 1}$. The
dimensionless Hookean spring force is given by
${\bm{\Vector{F}}}^c({\bm{\Vector{Q}}}_{\mu}) =
{\bm{\Vector{Q}}}_{\mu}$, while for FENE springs,
${\bm{\Vector{F}}}^c({\bm{\Vector{Q}}}_{\mu}) =
\dfrac{{\bm{\Vector{Q}}}_{\mu}}{1-{\vert\bm{\Vector{Q}}}_{\mu}\vert^2/b}$,
where $b = H q_0^2 /k_B T$ is the dimensionless finite extensibility
parameter, and $q_0$ is the dimensional maximum stretch of a spring.

The non-dimensional diffusion tensor $\bm{\Tensor{D}}_{\nu \mu}$ in
Eq.~(\ref{eq:sde}) is related to the non-dimensional hydrodynamic
interaction tensor ${\bm{\Tensor{\varOmega}}}$ through
\begin{equation}
\label{eq:Domega}
{\bm{\Tensor{D}}}_{\mu \nu} = \delta_{\mu \nu} \, \bm{\Tensor{\delta}} 
+ (1 - \delta_{\mu \nu}) \,
{\bm{\Tensor{\varOmega}}} (\bm{r}_\nu - \bm{r}_\mu)
\end{equation}
where $\bm{\Tensor{\delta}}$ and $\delta_{\mu \nu}$ represent a unit
tensor and a Kronecker delta, respectively, while
${\bm{\Tensor{\varOmega}}}$ represents the effect of the motion of a
bead $\mu$ on another bead $\nu$ through the disturbances carried by
the surrounding fluid. The hydrodynamic interaction tensor
${\bm{\Tensor{\varOmega}}}$ is assumed to be given by the
Rotne-Prager-Yamakawa (RPY) regularisation of the Oseen function
\begin{equation}
\label{eq:RPY}
{\bm{\Vector{\varOmega}}}({\bm{\Vector{r}}}) 
= {\varOmega}_1 \, \bm{\Tensor{\delta}} 
+ {\varOmega}_2 \frac{{\bm{\Vector{r}}} 
{\bm{\Vector{r}}}} {{{\Vector{r}}}^2}
\end{equation} 
where for $r \ge 2a$, the \emph{branch} $\mathcal{A}$ of the RPY
functions ${\varOmega}_1$ and ${\varOmega}_2$ is given by,
respectively,
\begin{equation}
\label{eq:b1}
{\varOmega}_1 = \frac{3a}{4r} \left(1 + \frac{2{a}^2}{3{r}^2} \right)  
\, \, \, \, \text{and} \, \, \, \, 
{\varOmega}_2 = \frac{3a}{4r} \left(1 - \frac{2{a}^2}{{r}^2} \right)
\end{equation}
while for $0 < r \le 2a$, the \emph{branch} $\mathcal{B}$ of the RPY
functions ${\varOmega}_1$ and ${\varOmega}_2$ is given by,
respectively,
\begin{equation}
\label{eq:b2}
{\varOmega}_1 = 1 - \frac{9}{32} \frac{r}{a} 
\, \, \, \, \text{and} \, \, \, \, 
{\varOmega}_2 = \frac{3}{32} \frac{r}{a}
\end{equation}
We introduce the notation of the two branches $\mathcal{A}$ and
$\mathcal{B}$ for facilitating subsequent discussion. The quantity $a$
has been introduced here as the non-dimensional radius of the bead as
an additional independent parameter. It is related to the
conventionally defined~\cite{thurston1967,bird} hydrodynamic
interaction parameter $h^*$ by $a = \sqrt{\pi} h^*$. As is well
known~\cite{beenakker}, the sum $\sum_{\mu} {\bm{\Tensor{D}}}_{\nu
  \mu} \cdot {\bm{\Vector{F}}}_{\mu} $ in Eq.~(\ref{eq:sde}) converges
slowly since ${\bm{\Tensor{D}}}_{\mu \nu}$ is long-ranged in nature,
scaling as $1/r$. The problem of slow convergence can be resolved
through the use of the Ewald sum, as discussed in greater detail
below. It is worth noting here that it is sufficient to evaluate
$\sum_{\mu} {\bm{\Tensor{D}}}_{\nu \mu} \cdot {\bm{\Vector{F}}}_{\mu}
$ in order to determine the time evolution of
${\bm{\Vector{r}}}_{\nu}(t)$. It is not necessary to know
$\bm{\Tensor{D}}_{\nu \mu}$ explicitly. Further, as will be seen
later, the evaluation of $\sum_{\mu} {\bm{\Tensor{B}}}_{\nu \mu} \cdot
{\Vector{\Delta \bm{W}}}_{\mu} $ using a Chebyshev polynomial
approximation for ${\bm{\Tensor{B}}}_{\nu \mu}$, also requires a
repeated evaluation of the Ewald sum.

\subsection{\label{sec:ewald-method} 
Evaluation of $\sum_{\mu} {\Tensor{D}}_{\nu \mu} 
\cdot {\Vector{F}}_{\mu} $ as an Ewald sum}

\begin{figure}[t]
\begin{center}
\includegraphics[width=0.48\textwidth]{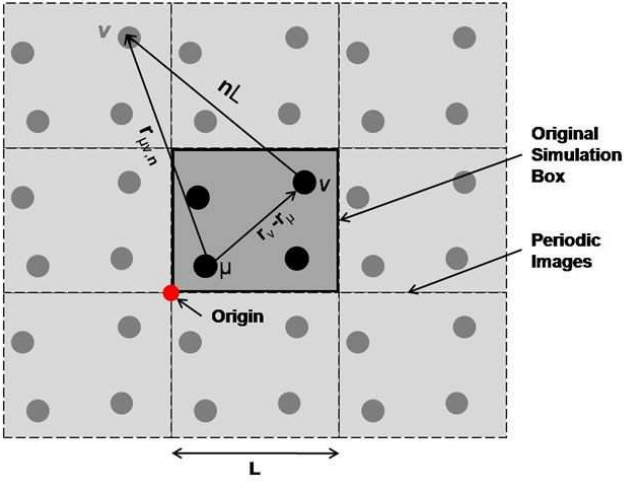}
\end{center}
\caption{(Color online) Periodic boundary conditions in 2-D:
  demonstration of the distance vector between two beads}
\label{pbc}
\end{figure}

Beenakker's~\cite{beenakker} representation of the sum $\sum_{\mu}
{\bm{\Tensor{D}}}_{\nu \mu} \cdot {\bm{\Vector{F}}}_{\mu}$ as an Ewald
sum for infinite periodic systems, using the RPY tensor to represent
hydrodynamic interactions, has the form
\begin{eqnarray}
\label{eq:DF}
\nonumber
&&
\sum_{\mu = 1}^{N} \bm{\Tensor{D}}_{\nu \mu} \cdot \bm{\Vector{F}}_{\mu}
= 
\left(1 - \frac{6 a \alpha}{\sqrt{\pi}} 
+ \frac{40 a^3\, {\alpha}^3}{3 \sqrt{\pi}}\right) \Vector{F}_{\nu}
\\
\nonumber
& + & 
{\sum_{\Vector{n} }}' \sum_{\mu = 1}^{N} 
\bm{\Tensor{M}}^{(1)} (\Vector{r}_{\nu\mu, \Vector{n}}) 
\cdot \Vector{F}_{\mu}  
+
\sum_{\Vector{k}\not= \bm{0}} \bm{\Tensor{M}}^{(2)}(\Vector{k}) \cdot
\\
\nonumber
& \bigg \{ &
\cos(\Vector{ k \cdot r_{\nu}}) 
\sum_{\mu = 1}^{N} \cos(\Vector{k \cdot r_{\mu}}) \Vector{F}_{\mu}
\\
& - &
\sin(\Vector{ k \cdot r_{\nu}}) 
\sum_{\mu = 1}^{N} \sin(\Vector{k \cdot r_{\mu}}) \Vector{F}_{\mu} 
\bigg \}
\end{eqnarray}
where the first and the second sums on the right hand side, both of
which converge rapidly, are carried out in real and reciprocal space,
respectively. The first term on the RHS is the correction due to
self-interactions and does not involve any summation. The parameter
$\alpha$ determines the manner in which the computational burden is
split between the two sums. The vector $\Vector{r}_{\nu\mu,
  \Vector{n}}$ is defined by $\Vector{r}_{\nu\mu, \Vector{n}} =
\Vector{r}_{\nu} - \Vector{r}_{\mu} + \Vector{n}L$, where $\Vector{n}
= (n_x, n_y, n_z)$ is the lattice vector with $n_x, n_y, n_z$ being
integer numbers (see Fig.~\ref{pbc}). The first summation on the RHS
of Eq.~(\ref{eq:DF}) is carried out in the original simulation box and
over all the neighbouring periodic images. The prime on the summation
indicates that the lattice vector $\Vector{n} = 0$ is omitted for $\nu
= \mu$. $\bm{\Tensor{M}}^{(1)}(\Vector{r})$ is a $3 \times 3$ matrix
(in real space), which depends on $a$ and $\alpha$, and
$\bm{\Tensor{M}}^{(2)}(\Vector{k})$ is also a $3 \times 3$ matrix (in
reciprocal space), which depends on $a,\alpha$ and the volume of the
simulation box $V$. The expressions for
$\bm{\Tensor{M}}^{(1)}(\Vector{r})$ and
$\bm{\Tensor{M}}^{(2)}(\Vector{k})$ are
\begin{eqnarray}
\label{eq:M1}
\nonumber
\bm{\Tensor{M}}^{(1)}(\Vector{r}) 
& = & 
\bigg [ \erfc(\alpha r) \, \left( \frac{3 a}{4 r} \, 
+ \, \frac{a^3}{2 r^3} \right)
\\
\nonumber
& + &
\frac{\exp{(-{\alpha}^2 r^2)}}{\sqrt{\pi}} 
\bigg ( 3 a \, {\alpha}^3 r^2 \, - \, \frac{9 a \alpha}{2} 
\, + \, 4 a^3 \, {\alpha}^7 r^4 
\\
\nonumber
& - &
20 a^3 \, {\alpha}^5 r^2 \, 
+ \, 14 a^3 \, {\alpha}^3 \, + \, \frac{a^3 \alpha}{r^2} 
\bigg ) \bigg ]
\, \bm{\Tensor{\delta}}
\\
\nonumber
& + &
\bigg [ \erfc(\alpha r) \, 
\left( \frac{3 a}{4 r} \, - \, \frac{3 a^3}{2 r^3} \right)
\\
\nonumber
& + &
\frac{\exp{(-{\alpha}^2 r^2)}}{\sqrt{\pi}}\, 
\bigg ( \frac{3 a \alpha}{2} \, - \, 3 a \, {\alpha}^3 r^2  
- 4 a^3 \, {\alpha}^7 r^4 
\\
& + &
16 a^3 \, {\alpha}^5 r^2 \, 
- \, 2 a^3 \, {\alpha}^3 \, - \, \frac{3 a^3 \alpha}{r^2} 
\bigg ) \bigg ]
\, \hat{\Vector{r}} \hat{\Vector{r}}
\end{eqnarray}
with $\erfc$ denoting the complementary error function, and
\begin{eqnarray}
\label{eq:M2}
\nonumber
\bm{\Tensor{M}}^{(2)}(\Vector{k}) 
& = &
\left(a \, - \, \frac{a^3 k^2}{3} \right) \,\, 
\left(1 \, + \, \frac{k^2}{4 \, {\alpha}^2} \, 
+ \, \frac{k^4}{8 {\alpha}^4} \right) 
\\
&&
\left(\frac{6 \pi}{k^2 V} \right)
\exp \left(\frac{-k^2}{4 {\alpha}^2} \right)
\, \left(\bm{\Tensor{\delta}} \, - \, \hat{\Vector{k}} 
\, \hat{\Vector{k}} \right) 
\end{eqnarray}

The second summation in Eq.~(\ref{eq:DF}) (denoted here as the
reciprocal space sum) is carried out over lattice vectors $\Vector{k}
= 2\pi \Vector{n}/L$. In Eq.~(\ref{eq:M1}), $r$ and $\hat{\Vector{r}}$
are the magnitude and unit vector, respectively, in the direction of
$\Vector{r}$. In Eq.~(\ref{eq:M2}), $k$ and $\hat{\Vector{k}}$ are the
magnitude and unit vector, respectively, corresponding to
$\Vector{k}$.

\subsection{\label{subsec:ewald-method} 
Modification of the Ewald sum to account 
for overlapping beads}

As pointed out earlier, the derivation of the Ewald sum by
\citet{beenakker} is valid \emph{only} for the {branch} $\mathcal{A}$
of the RPY functions ${\varOmega}_1$ and ${\varOmega}_2$
(Eq.~(\ref{eq:b1})), which forbids its use for the case of overlapping
beads ($r < 2a$). The original expression consequently cannot be used
for the simulation of $\theta$ solvents by neglecting excluded-volume
interactions, as in this case beads on the same or on different
chains are highly prone to overlap with each other. The Ewald sum has
been modified here to account for such situations.

Starting from a given bead $\nu$, we consider all those beads that
have distances less than $2 a$ from it, including bead $\nu$
itself. By a proper re-labeling, we can assume that these are the
beads $\mu = 1, \ldots, N^*$. The number of non-overlapping particles
is thus $N - N^*$. As the correction needs to be carried out only in
the real space sum, the first summation on the RHS of
Eq.~(\ref{eq:DF}) $\left({\sum^{\prime}_{\Vector{n}}} \sum_{\mu =
    1}^{N} \bm{\Tensor{M}}^{(1)}(\Vector{r}_{\nu\mu, \Vector{n}})
  \cdot \Vector{F}_{\mu}\right)$ is replaced by
\begin{eqnarray}
\label{eq:cor1}
\nonumber
&& 
\sum_{\substack{ {\mu = 1}\\\mu \neq \nu}}^{N^*}  
\bm{\Tensor{M_B}}^{(1)}(\Vector{r}_{\nu\mu, \Vector{n = 0}}) 
\cdot \Vector{F}_{\mu} 
\, + \, 
\sum_{\Vector{n \neq \bm{0}}} \sum_{\mu = 1}^{N^*} 
\bm{\Tensor{M}}^{(1)}(\Vector{r}_{\nu\mu, \Vector{n}}) 
\cdot \Vector{F}_{\mu}
\\
& & + \,
{\sum_{\Vector{n}}} \sum_{\mu = N^* + 1}^{N } 
\bm{\Tensor{M}}^{(1)}(\Vector{r}_{\nu\mu, \Vector{n}})
\cdot \Vector{F}_{\mu}  
\end{eqnarray}
where the first summation of Eq.~(\ref{eq:cor1}) is carried out only
over overlapping particles in the original simulation box ($\Vector{n
  = 0}$). Similar to Beenakker's~\cite{beenakker} derivation of the
$\bm{\Tensor{M}}^{(1)}$ matrix, the matrix $\bm{\Tensor{M_B}}^{(1)}$
is derived here based on the {branch} $\mathcal{B}$ of the RPY tensor
given by Eq.~(\ref{eq:b2}). The second summation is carried out over
the periodic images of the overlapping particles (whose distances are
more than $2a$), and the third summation is similar to that given in
Eq.~(\ref{eq:DF}) but here it is carried out only over the
non-overlapping particles. Note that the second sum is not carried out
in the original box.  In order to make this sum extend over the
original box and periodic images, a term is added and subtracted as
follows
\begin{eqnarray}
\label{eq:cor2}
\nonumber
&&
\sum_{\substack{ {\mu = 1} \\ \mu \neq \nu}}^{N^*}  
\bm{\Tensor{M_B}}^{(1)}(\Vector{r}_{\nu\mu, \Vector{n = 0}}) \cdot \Vector{F}_{\mu} 
\, + \, \sum_{\Vector{n \neq 0}} \sum_{\mu = 1}^{N^*} 
\bm{\Tensor{M}}^{(1)}(\Vector{r}_{\nu\mu, \Vector{n}}) \cdot \Vector{F}_{\mu} 
\\
\nonumber
& + &
\sum_{\substack{ {\mu = 1}\\\mu \neq \nu}}^{N^*}  
\bm{\Tensor{M}}^{(1)}(\Vector{r}_{\nu\mu, \Vector{n=0}}) \cdot \Vector{F}_{\mu}   
- \, \sum_{\substack{ {\mu = 1}\\\mu \neq \nu}}^{N^*} 
\bm{\Tensor{M}}^{(1)}(\Vector{r}_{\nu\mu, \Vector{n=0}}) \cdot \Vector{F}_{\mu} 
\\
& + & 
{\sum_{\Vector{n}}} \sum_{\mu = N^* + 1}^{N } 
\bm{\Tensor{M}}^{(1)}(\Vector{r}_{\nu\mu, \Vector{n}}) \cdot \Vector{F}_{\mu}  
\end{eqnarray} 
The second, third and fifth summations of the above equation together
represent the original real space sum in
Eq.~(\ref{eq:DF}). Eq.~(\ref{eq:cor2}) can consequently be rearranged
as
\begin{eqnarray}
\label{eq:cor3}
\nonumber
&&
\left({\sum_{\Vector{n}}}' \sum_{\mu = 1}^{N} 
\bm{\Tensor{M}}^{(1)}(\Vector{r}_{\nu\mu, \Vector{n}}) 
\cdot \Vector{F}_{\mu}\right)
\\
& + &
\sum_{\substack{ {\mu = 1}\\\mu \neq \nu}}^{N^*} 
\left[\bm{\Tensor{M_B}}^{(1)}(\Vector{r}_{\nu\mu, \Vector{n = 0}}) 
- \, \bm{\Tensor{M}}^{(1)}(\Vector{r}_{\nu\mu, \Vector{n = 0}}) 
\right] \cdot \Vector{F}_{\mu}
\end{eqnarray}
where the second summation is carried over all overlapping particles
in the original box. Denoting the second term in Eq.~(\ref{eq:cor3})
by $\bm{\Tensor{M}}^*$, it is straightforward to show that
\begin{eqnarray}
\label{eq:cor5}
\nonumber
\bm{\Tensor{M}}^* (\Vector{x}) 
& =  &
\left[ 1 - \frac{1}{2 x^3}{\left(\frac{3 x^2}{4} + 1 \right)^2}
\right] \, \bm{\delta}
\\
& + &
\left[ \frac{1}{2 x^3}{\left(\frac{3 x^2}{4} - 1 \right)^2}
\right] \, \hat{\Vector{x}} \, \hat{\Vector{x}}
\end{eqnarray}
where $\Vector{x} = \Vector{r}_{(\nu\mu, \Vector{n = 0})} /a$ and
$\hat{\Vector{x}}$ is the unit vector in the direction of
$\Vector{r}_{(\nu\mu, \Vector{n = 0})}$. The modified form of the
Ewald sum that is valid for arbitrary inter-particle distance is
consequently
\begin{eqnarray}
\label{eq:DFmodified}
\nonumber
&&
\sum_{\mu = 1}^{N} \bm{\Tensor{D}}_{\nu \mu} \cdot \bm{\Vector{F}}_{\mu}
= 
\left(1 - \frac{6 a \alpha}{\sqrt{\pi}} 
+ \frac{40 a^3\, {\alpha}^3}{3 \sqrt{\pi}}\right) \Vector{F}_{\nu}
\\
\nonumber
& + & 
{\sum_{\Vector{n} }}' \sum_{\mu = 1}^{N} 
\bm{\Tensor{M}}^{(1)} (\Vector{r}_{\nu\mu, \Vector{n}}) 
\cdot \Vector{F}_{\mu}  
+
\sum_{\Vector{k}\not= \bm{0}} \bm{\Tensor{M}}^{(2)}(\Vector{k}) \cdot
\\
\nonumber
& \bigg \{ &
\cos(\Vector{ k \cdot r_{\nu}}) 
\sum_{\mu = 1}^{N} \cos(\Vector{k \cdot r_{\mu}}) \Vector{F}_{\mu}
\\
\nonumber
& - &
\sin(\Vector{ k \cdot r_{\nu}}) 
\sum_{\mu = 1}^{N} \sin(\Vector{k \cdot r_{\mu}}) \Vector{F}_{\mu} 
\bigg \}
\\
& + & \sum_{\substack{ {\mu = 1}\\\mu \neq \nu}}^{N^*}
\bm{\Tensor{M}}^{*}(\Vector{r}_{\nu\mu, \Vector{n = 0}})
\cdot \Vector{F}_{\mu}
\end{eqnarray}

\subsection{\label{sec:implement-ewald} 
Implementation of the Ewald sum}

As discussed earlier, the real space sum is carried out over all the
periodic images while the reciprocal space sum is performed only in
the original simulation box. There are three parameters which control
the accuracy of both the real and reciprocal space sums: $n_{max}$,
an integer which defines the range of the real space sum (governed by
the number of periodic images, see Fig.~\ref{pbc}), $k_{max}$, an
integer that defines the summation range in reciprocal space and the
Ewald parameter $\alpha$. These three parameters are related to each
other from the point of view of accuracy and speed. A large value
of $\alpha$ makes the real space sum converge faster (since a smaller
value of $n_{max}$ is required). However, this leads to the reciprocal
space sum requiring a larger number of wavevectors $k_{max}$. On the
other hand, a small value of $\alpha$ implies an expensive real space
sum but a cheaper reciprocal space sum. The optimal choice of these
parameters has been discussed previously by \citet{fincham} in the
context of electrostatic interactions. Here, a similar study is
performed for hydrodynamics interactions.

\subsubsection{\label{subsec:choice} Choice of Ewald parameters}

At fixed monomer bulk concentration $c$, the box size increases as
$N^{1/3}$.  As can be seen from Eqs.~(\ref{eq:DF}) and (\ref{eq:M1}),
the convergence of the real space sum depends on the complementary
error function $\erfc(\alpha r)$, where $r$ is the distance between a
pair of beads. In practice, the sum is evaluated only for $r \le r_{\text{c}}$,
where $r_{\text{c}}$ denotes a \emph{cutoff} radius. The value of $\alpha$ is
chosen such that $\erfc(\alpha r_{\text{c}})$ is small.  At large values of the
argument, $\erfc(\alpha r_{\text{c}})$ behaves as $\exp(-{\alpha}^2
{r_{\text{c}}}^2)$. If we specify $M$ such that $\exp(-M^2)$ is very small,
then
\begin{equation}
\label{eq:rparameter}
{\alpha}^2 {r_{\text{c}}}^2 = M^2 \, \, \, \,   
\text{or} \, \, \, \,  \alpha = M/r_{\text{c}}
\end{equation} 
Similarly the rate of convergence of the reciprocal space sum is
controlled by the factor $\exp(-k^2/4 {\alpha}^2)$. If it is
required~\cite{fincham} that the accuracy of the real space sum is
roughly equal to that of the reciprocal space sum at the reciprocal
space cutoff, $K$, then using Eq.~(\ref{eq:rparameter}) we find
\begin{equation}
\label{eq:kparameter}
M^2 = K^2 / (4 {\alpha}^2) \, \, \, \,   
\text{or} \, \, \, \,  
K = 2 \alpha M = 2 M^2 /r_{\text{c}}
\end{equation}
These relations allow us to specify $\alpha$ and $K$ for given values
of $M$ and $r_{\text{c}}$, while the latter parameters control the accuracy and
speed of the algorithm, as discussed subsequently.

\subsubsection{\label{subsec:r-imple} 
The real space and reciprocal space sums}

Locating all pairs of beads which are separated by less than the
cutoff distance $r_{\text{c}}$ is the first step in evaluating the real space
sum. A naive all-pairs neighbour search results in $O(N^2)$
performance and therefore the link-cell method, which is a cell-based
neighbour search method, is used to improve the
performance~\cite{HGE,griebel,vectorMD}. The calculation of the
infinite real space sum is thus reduced to the calculation of the sum
locally over only a small number of neighbouring beads. Here, the
neighbour search is implemented with cells of side $r_{\text{c}}/5$.
 
The reciprocal space sum is more straightforward to implement. The
major effort is expended in the evaluation of terms of the form
$\exp(i \bf{k \cdot {r}}_\mu)$. The method adopted here precomputes
the components of these factors by recursion and stores
them~\cite{fincham}. This avoids calling the complex exponential
function repeatedly. However, it involves a substantial amount of
computer memory.

%
\section{\label{sec:opt-DF} Optimisation of the
evaluation of $\sum_{\mu} {\Tensor{D}}_{\nu \mu}
\cdot {\Vector{F}}_{\mu} $}

\begin{figure*}[t]
\centering
\begin{tabular}{cc}
\includegraphics[width=0.48\textwidth]{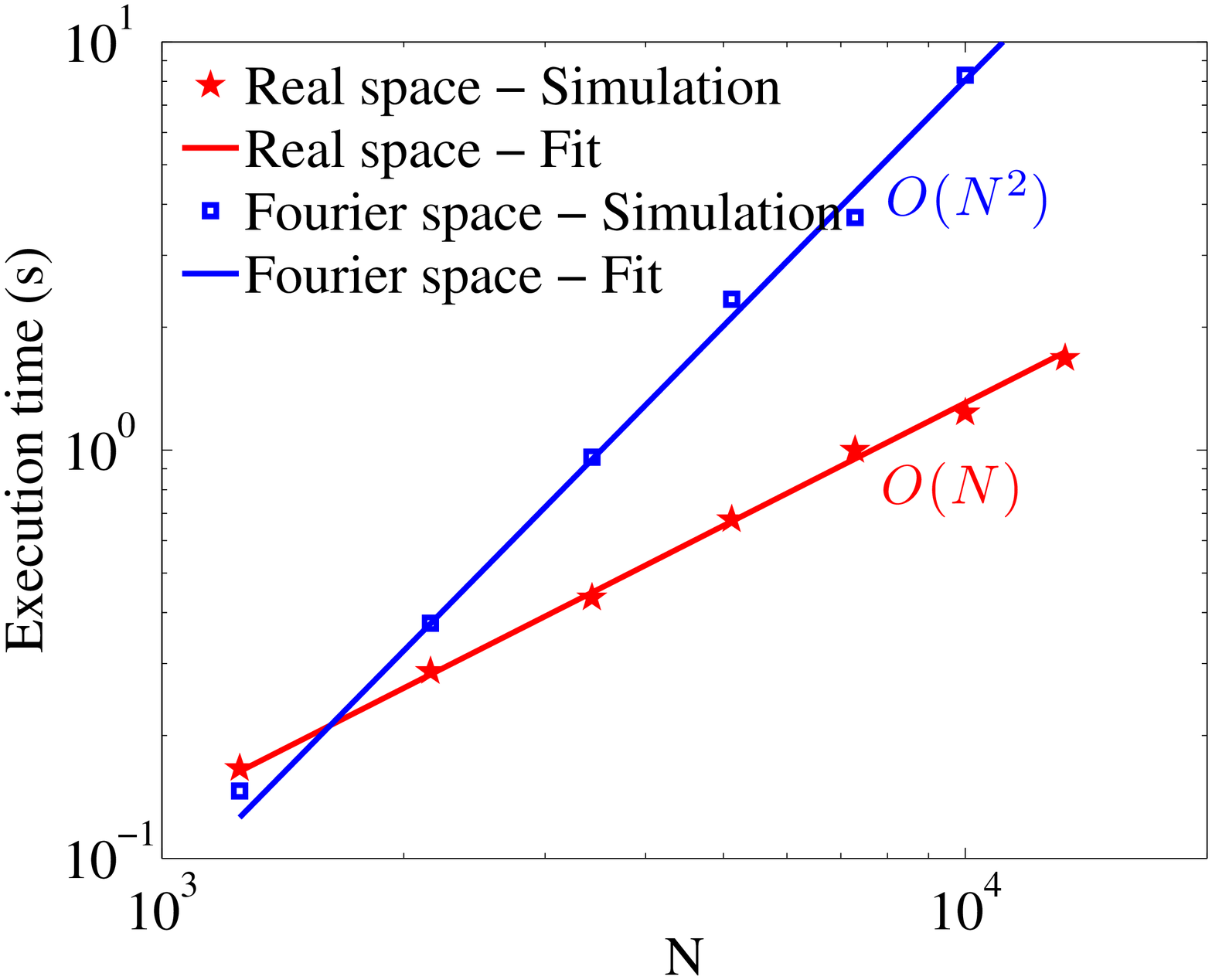} &
\includegraphics[width=0.48\textwidth]{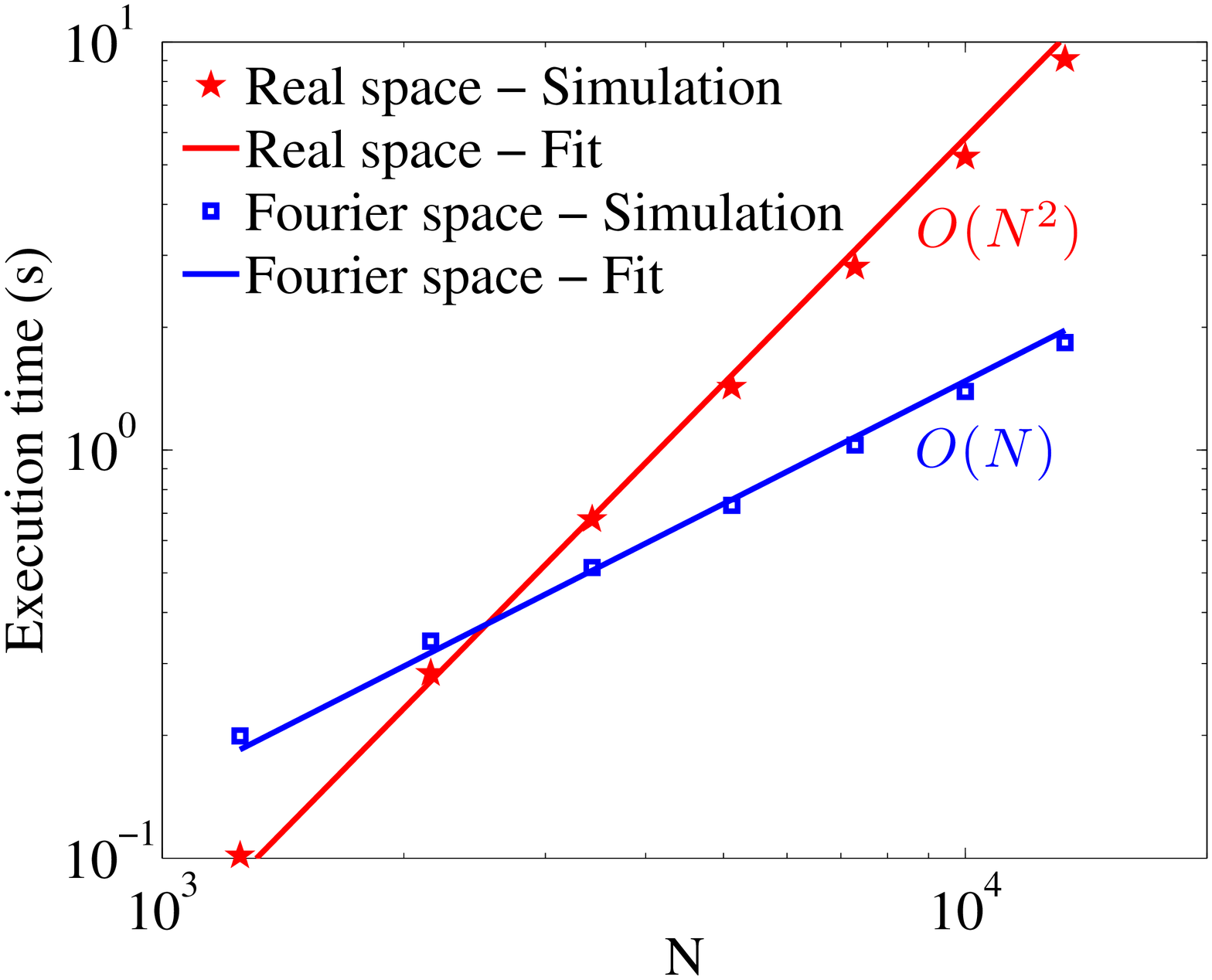} \\
(a) & (b)  \\
\end{tabular}
\caption{(Color online) \small \label{num_vs_ana_opt} Execution time
  scaling for the real space and the Fourier space sums for: (a)
  Constant $r_{\text{c}}$ (b) Constant $L/r_{\text{c}}$}
\end{figure*}

The Ewald parameter $\alpha$, which splits the computational burden
between the real space sum and the reciprocal space sum, is related to
the real space cutoff $r_{\text{c}}$ by Eq.~(\ref{eq:rparameter}). The aim of
optimisation is to minimise the total execution time (which is the sum
of the real space execution time, $T_R$ and the reciprocal space
execution time, $T_F$), with respect to the real space cutoff
$r_{\text{c}}$. Following \citet{fincham}, the execution time $T_R$ is
calculated as follows. A sphere of cutoff radius $r_{\text{c}}$ contains on
average $N_{\text{r}_{\text{c}}} = \dfrac{4 \pi}{3} \, r_{\text{c}}^3 \, c$
beads. Each bead interacts with the $N_{\text{r}_{\text{c}}}$ beads
that surround it. Since for symmetry reasons, each pair interaction
needs to be considered only once, the execution time is
\begin{equation}    
\label{eq:TRDF}
T_R = \frac{1}{2} \, A_R  \, N \, 
\frac{4 \pi}{3} \, r_{\text{c}}^3 \, c \, t_r
\end{equation}
where $A_R$ is a constant that depends on the code architecture and
$t_r$ is the execution time to evaluate one interaction, which is
found to be $0.15 \, \mu s$ when the BD code is run on a 156 SGI Altix
XE 320 cluster. Eq.~(\ref{eq:TRDF}) is fit to data obtained by running
simulations for a range of parameters on a 156 SGI Altix XE 320
cluster, and the fitted parameter $A_R$ is then found to be $5$.

The execution time $T_F$ is evaluated as follows. Within the cutoff
$K$, the volume of the reciprocal space is $\dfrac{4 \pi}{3} K^3 =
\dfrac{4 \pi}{3} \dfrac{8 M^6}{r_{\text{c}}^3}$ (the latter follows from
Eq.~(\ref{eq:kparameter})). The reciprocal space points are defined by
$k = \dfrac{2 \pi}{L} (l, m, n)$ where $l, m, n$ are integers and $L$
is the simulation box size. The volume of reciprocal space per point
is, thus, ${(2 \pi/L)}^3$, and $\dfrac{4 \pi}{3} \dfrac{8 M^6}{r_{\text{c}}^3}
\dfrac{L^3}{8 {\pi}^3}$ is the number of points in the cutoff
sphere. Using $L^3 = N/c$ to highlight the $N$ dependence at fixed
concentration $c$, the number of reciprocal space points in the cutoff
sphere becomes $\dfrac{4 \pi}{3} \dfrac{M^6}{{\pi}^3} \dfrac{N}{c
  r_{\text{c}}^3}$. It is worth pointing out that for fixed cutoff radius, the
number of $k$-space points increases as $N$, because the concentration
of points in reciprocal space increases with system size. Further,
inversion symmetry of reciprocal space halves the number of reciprocal
space points mentioned above. A sum over the $N$ beads must be
performed for each $k$-space point, so the execution time is
\begin{equation}    
\label{eq:TFDF}
T_F = A_F \, \frac{1}{2} \, \frac{4 \pi}{3} \, \frac{M^6}{{\pi}^3} 
\, \frac{N^2}{c r_{\text{c}}^3} \, t_f
\end{equation}  
where $A_F$ is a code architecture constant and $t_f$ is the execution
time to evaluate one term in the sum, which is found to be $0.33 \,
\mu s$. As in the real space instance, Eq.~(\ref{eq:TFDF}) is fitted
to simulation data to obtain $A_F = 0.19$. The total execution time is
consequently
\begin{eqnarray}
\label{eq:TDF}
\nonumber
T & = & T_R + T_F 
\\
& = & \frac{1}{2} \, \frac{4 \pi}{3} \, \left[A_R \, N \, 
r_{\text{c}}^3 \, c \, t_r + A_F \frac{M^6}{{\pi}^3} \, 
\frac{N^2}{c r_{\text{c}}^3} \, t_f \right]
\end{eqnarray}

\begin{figure*}[t]
\centering
\begin{tabular}{cc}
\includegraphics[width=0.48\textwidth]{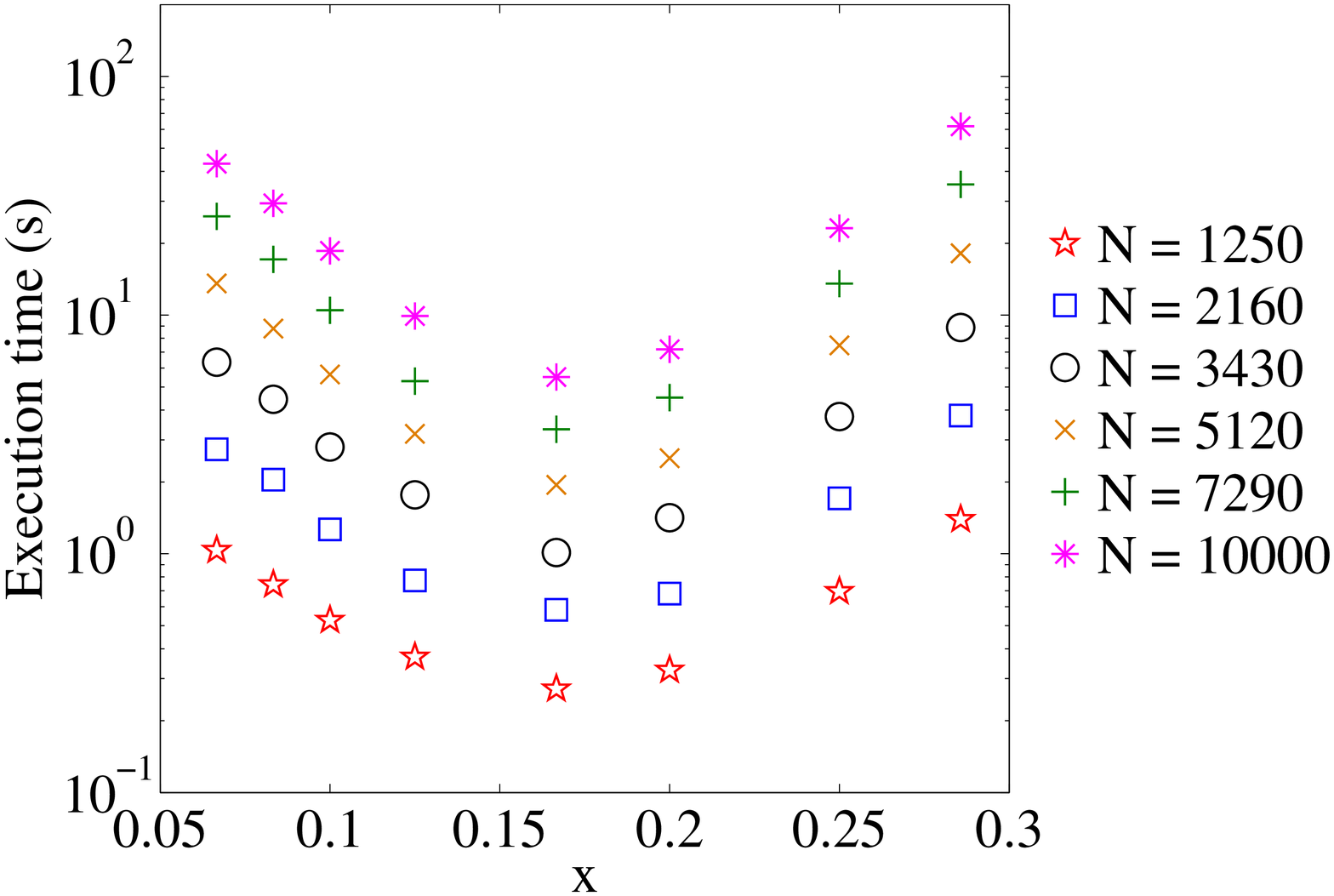} &
\includegraphics[width=0.48\textwidth]{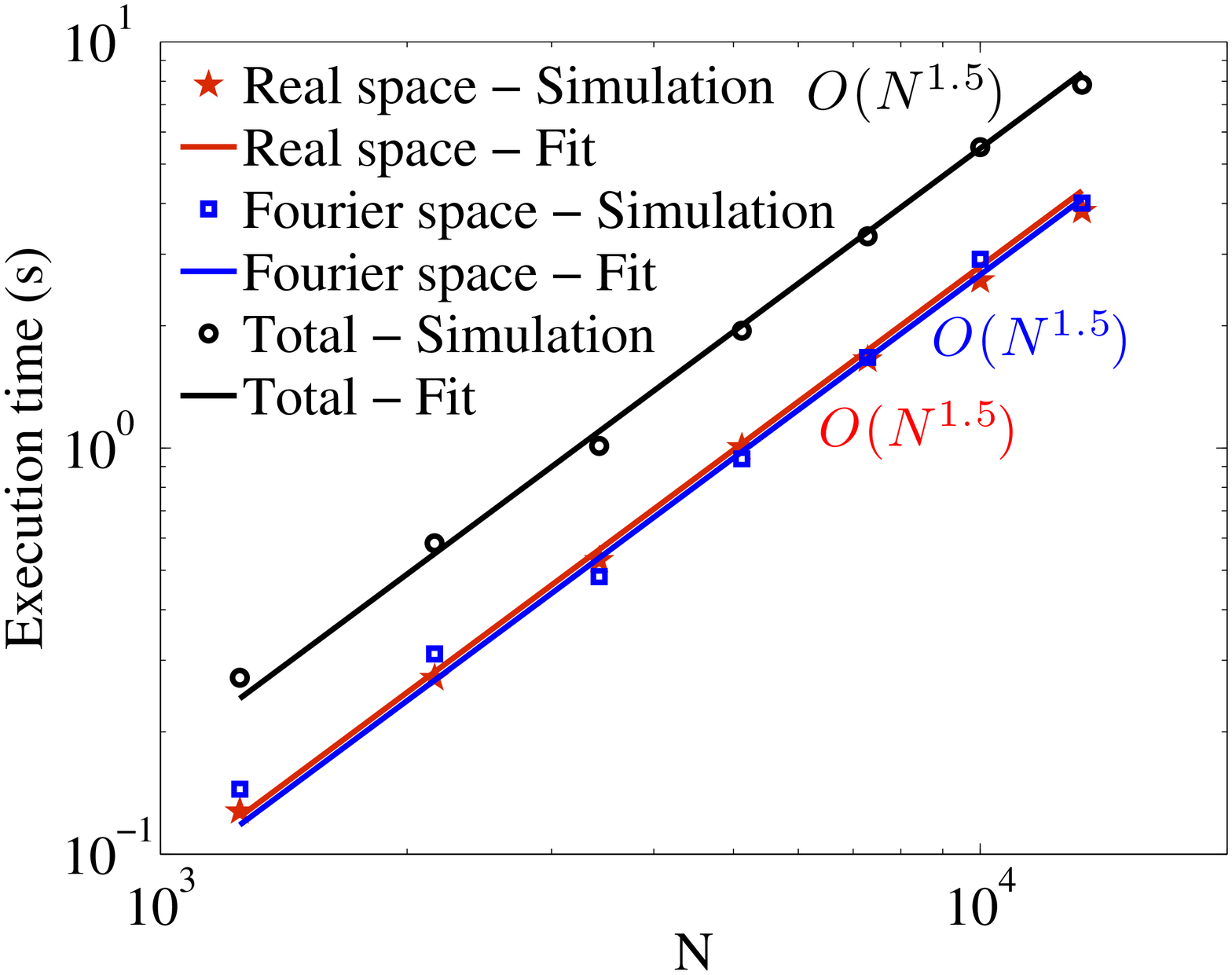} \\
(a) & (b)  \\
\end{tabular}
\caption{(Color online) \small \label{opt-scaling} (a) Total execution
  time vs. exponent $x$ for various $N$ (b) Power law scaling for
  $T_R$, $T_F$ and $T_{\text{opt}}$ at
  ${\left(r_{\text{c}}^{\text{E}}\right)}_{\text{opt}}$}
\end{figure*}

Equation~(\ref{eq:TDF}) shows that, for fixed $M$ and $r_{\text{c}}$,
$T_R$ varies as $N$, but $T_F$ varies as $N^2$, because of the
increasing concentration of points in reciprocal space. This behaviour
is demonstrated for $c = 4.44 \, c^\star$ and $r_{\text{c}} = 10$ in
Fig.~\ref{num_vs_ana_opt}~(a) for the simulation data (symbols), which
agrees with the expressions given in Eqs.~(\ref{eq:TRDF})
and~(\ref{eq:TFDF}) (solid lines). Here $c^\star$ is the overlap
concentration defined by $N_b/\left[\dfrac{4 \pi}{3}
  (R_g^{0})^3\right]$, where $R_g^0$ is the radius of gyration of a
polymer chain in the dilute limit. To increase the value of $N$, we
fix the value of beads per chain at $N_b = 10$ and increase the number
of chains $N_c$. Conversely, if $r_{\text{c}}$ is increased as the
system size increases in such a way that $r_{\text{c}}/L$ is constant
(as in the approach adopted by \citet{stoltz}), then since $c =
N/L^3$, $T_R$ varies as $N^2$ but $T_F$ varies as
$N$. Figure~\ref{num_vs_ana_opt}~(b) displays this behaviour for $N_b =
10$, $c = 4.44 \, c^\star$, $M = 3.3$ and $L/r_{\text{c}} = 3$. Once
again the simulation data is seen to match the expressions given in
Eqs.~(\ref{eq:TRDF}) and~(\ref{eq:TFDF}). This suggests that by
appropriate choice of parameters it may be possible to achieve better
than $N^2$ behaviour in the total time $T$. For a given accuracy, the
only free parameter is $r_{\text{c}}$, since this determines $\alpha$
and hence $K$ by Eqs.~(\ref{eq:rparameter}) and
(\ref{eq:kparameter}). To find the value of $r_{\text{c}}$ which
minimises the total execution time, we set $dT/dr_{\text{c}} =
0$. This leads to
\begin{equation}
\label{eq:ropt_opt}
{\left(r_{\text{c}}^{\text{E}}\right)}_{\text{opt}} 
= \frac{M}{\sqrt{\pi}} \, \, 
{\left(\frac{A_F \, t_f}{A_R \, t_r}\right)}^{1/6} \, \,
\frac{N^{1/6}}{c^{1/3}}
\end{equation} 
Thus the optimal choice of the cutoff radius
${\left(r_{\text{c}}^{\text{E}}\right)}_{\text{opt}}$ increases slowly
($1/6^{\text{th}}$ power) with $N$. The validity of
Eq.~(\ref{eq:ropt_opt}) has been verified by carrying out
simulations. Assuming that the quantity $N^{1/6}$ in
Eq.~(\ref{eq:ropt_opt}) is replaced by $N^x$, various values of the
exponent $x$ are selected in place of the exponent $1/6$, and the
total execution time for a given value of $N$ is
estimated. Fig.~\ref{opt-scaling}~(a) shows the total execution time
as a function of the exponent $x$ and it is clear that the minimum
execution time is achieved when $x = 1/6$ as given by
Eq.~(\ref{eq:ropt_opt}), for all $N$.

Substituting ${\left(r_{\text{c}}^{\text{E}}\right)}_{\text{opt}}$ in
Eq.~(\ref{eq:TDF}) we find for the optimal time
\begin{equation}
\label{eq:Topt}
T_{\text{opt}} = 2 T_R = 2 T_F = 
\frac{4 \pi}{3} \, \, N^{1.5} \, \, 
\frac{M^3}{{\pi}^{1.5}} \, \, \sqrt{A_R \, t_r \, A_F \, t_f}
\end{equation}
Thus, when the total time is optimised, it is equally divided between
the real and reciprocal space parts of the calculation. This is
verified in Fig.~\ref{opt-scaling}~(b), which displays plots of $T_R$,
$T_F$ and $T$ as a function of $N$, at $x = 1/6$ and for $N_b = 10$,
$c = 4.44 c^\star$ and $M = 3.3$. Symbols indicate simulation data and
solid lines correspond to Eq.~(\ref{eq:Topt}) with the appropriate
values for the various parameters. Eq.~(\ref{eq:Topt}) also indicates
that the real space, reciprocal space and total time scale as
$N^{1.5}$. Simulation results shown in Fig.~\ref{opt-scaling}~(b)
substantiate this prediction. These results are similar to those
obtained by \citet{fincham} in the context of electrostatic
interactions.

\section{\label{sec:chebyshev} Decomposition
of the diffusion tensor}

In component form, the decomposition rule (Eq.~(\ref{decomp})) for
obtaining the block matrix $\bcal{B}$ can be expressed as follows,
\begin{equation}
\sum_{\beta = 1}^{N} \sum_{q=1}^{3} \mathcal{B}_{\nu \beta}^{rq} \, 
\mathcal{B}_{\mu \beta}^{sq}   =  \mathcal{D}_{\nu \mu}^{rs} 
\label{decomp2}
\end{equation}
where $\{\nu, \beta, \mu = 1,\ldots,N\}$, $\{r,q,s = 1,2,3\}$, and
$\mathcal{D}_{\nu \mu}^{rs}$ is the `$r\!s$'$^\text{th}$ Cartesian
component of the tensor $ \bm{\Tensor{D}}_{\nu\mu}$. The matrix
$\bcal{B}$ is not unique. Assuming that $\bcal{B}$ is a lower (or
upper) triangular matrix leads to its calculation using a Cholesky
decomposition of $\bcal{D}$, which as mentioned earlier, requires
$O({N^3})$ operations. Fixman's~\cite{fixman} approach achieves an
attenuation of this CPU intensity by recognising that (i) it is
sufficient to find $\bcal{B}$ approximately, and (ii) the individual
columns of the matrix $\bcal{B}$ are in themselves not of much
interest, only the vector $d \bcal{S} = \bcal{B} \cdot \Delta
\bcal{W}$ is required, where $\Delta \bcal{W}$ is a vector consisting
of the $3N$ Gaussian noise coordinates ${\Delta W}_\mu^s$, with $\mu =
1,\ldots,N$, and $s = 1,2,3$. By assuming that $\bcal{B} =
\sqrt{\bcal{D}}$, and using a Chebyshev polynomial approximation for
the square root function, Fixman showed
that~\cite{fixman,jendrejack,kroger,prabhakarthesis}
\begin{gather}
d \bcal{S} = \sqrt{\bcal{D}}  \cdot \Delta \bcal{W} \approx
\sum_{p = 0}^{N_\text{Ch} - 1} c_p \, \bcal{V}_p - \frac{c_0}{2} \,
\Delta \bcal{W} \label{cheb1}
\end{gather}
where $N_{\text{Ch}}$ is the number of terms used in the Chebyshev
polynomial approximation, $c_p$ are the Chebyshev coefficients, and
the vectors $ \bcal{V}_p$ are defined by the recurrence relations
\begin{gather}
 \bcal{V}_p = 2 \bcal{Y} \cdot  \bcal{V}_{p -1}-  \bcal{V}_{p-2} 
 \, ;  \quad p \ge 2
\label{cheb2}
\end{gather}
with $ \bcal{V}_0 = \Delta \bcal{W}$ and $ \bcal{V}_1 = \bcal{Y} \cdot
\bcal{V}_0$. The linear mapping
\begin{gather}
\bcal{Y}  = \left(\frac{2}{d_\textrm{max} -
d_\textrm{min}}\right) \,\bcal{D} -
\left( \frac{ d_\textrm{max} + d_\textrm{min}}
{d_\textrm{max} - d_\textrm{min}} \right) \,
\bcal{I}
\label{cheb3}
\end{gather}
(where $\bcal{I}$ denotes the $3N \times 3N$-dimensional identity
matrix), ensures that the eigenvalues of $\bcal{Y}$ lie in the domain
$[-1, 1]$ when the eigenvalues of $\bcal{D}$ are within
$[d_\textrm{min}, d_\textrm{max}]$. This is essential for the validity
of the Chebyshev approximation.

It is clear that for a given $N_\text{Ch}$, the cost of the direct
calculation of the $3 N$-dimensional vector $d \bcal{S}$, without the
intermediate calculation of $\bcal{B}$, is proportional to
$N_{\text{Ch}} \, \times$ [the cost of evaluating $ \bcal{V}_p$]. The
number of terms $N_{\text{Ch}}$ that are required is determined by the
desired accuracy in the estimation of the square root. The choice of
$N_{\text{Ch}}$ is also affected by the necessity of ensuring that the
bounds $d_\textrm{max} $ and $d_\textrm{min}$ satisfy the following
constraints relative to the maximum ($\lambda_{\text{max}}$) and
minimum ($\lambda_{\text{min}}$) eigenvalues of $\bcal{D}$, namely,
$d_\textrm{max} \ge \lambda_{\text{max}} $ and $ d_\textrm{min}
\le\lambda_{\text{min}} $.

The CPU cost involved in adopting Fixman's procedure for dilute
polymer solutions have been discussed in depth in
Refs.~\onlinecite{jendrejack,kroger,prabhakarthesis}. Here, we briefly
summarise the main conclusions: (i) The cost of evaluating
$\bcal{V}_p$ is simply $O({N^2})$. (ii) The number of terms in the
Chebyshev approximation is determined using the
expression~\cite{kroger,prabhakar}
\begin{gather}
N_\text{Ch} = \mathrm{nint} \left[
\left(\frac{\lambda_{\text{max}}}%
{\lambda_{\text{min}} }\right)^{\frac{1}{2}} \right] 
+ 1 \label{nint}
\end{gather}
where $\mathrm{nint}$ is the nearest integer function. The use of
Eq.~(\ref{nint}) is motivated by the finding~\cite{fixman,jendrejack}
that the value of $N_\text{Ch}$ required to keep the error in the
estimation of the square root below a fixed tolerance, scales as
$({\lambda_{\text{max}}}/{\lambda_{\text{min}}
})^{\frac{1}{2}}$. (iii) The limiting eigenvalues
$\lambda_{\text{max}} $ and $\lambda_{\text{min}} $ can be calculated
exactly in $O({N^2})$ operations using standard software
packages---this procedure was adopted in Ref.~\onlinecite{jendrejack}
with the package ARPACK. On the other hand, \citet{kroger} and
\citet{prabhakar} avoid explicit evaluation of the eigenvalues, but
instead obtain approximate estimates for $\lambda_{\text{max}} $ and
$\lambda_{\text{min}} $. In particular, \citet{prabhakar} use the
following expressions based on a suggestion by \citet{fixman}
\begin{eqnarray}
\label{eq:lmax-lmin-dilute}
\nonumber
\lambda_{\text{max}}^{\text{Fixman}} 
& = & \frac{1}{3 N} \, \, (\bcal{U}^+ \cdot \bcal{D}
\cdot \bcal{U}^+)
\hspace{0.5cm}
\text{and} 
\\
\lambda_{\text{min}}^{\text{Fixman}}
& = & \frac{1}{3 N} \, \, (\bcal{U}^- \cdot \bcal{D}
\cdot \bcal{U}^-)
\end{eqnarray}
where $\bcal{U}^+$ is a $3 N$-dimensional vector, all of
whose elements are equal to 1 and $\bcal{U}^-$ is a $3
N$-dimensional vector with alternating $1\text{'}$s and
$-1\text{'}$s. Further, the bounds $d_\textrm{max} = 2
\lambda_{\text{max}}^{\text{Fixman}} $ and $d_\textrm{min} =
0.5\lambda_{\text{min}}^{\text{Fixman}} $ were chosen to satisfy the
conditions on the magnitudes of $d_\textrm{max}$ and $d_\textrm{min}$
relative to the maximum and minimum eigenvalues. Since for dilute
polymer solutions, $({\lambda_{\text{max}}}/{\lambda_{\text{min}} })
\sim N^{0.5}$ and consequently $N_\text{Ch} \sim N^{1/4}$, the
CPU-time requirement for the calculation of $d \bcal{S}$ in Fixman's
method scales as $N_\text{Ch} N^2 \sim N^{9/4}$.

In the case of semidilute polymer solutions, since determining
$\bcal{V}_p$ requires the recursive evaluation of the product of a
linear transformation of the diffusion tensor with various
$3N$-dimensional vectors (see Eq.~(\ref{cheb2})), the Ewald sum can be
used for its evaluation, with the force vector $\textbf{F}_{\mu}$ in
Eq.~(\ref{eq:DF}) replaced by the relevant vector in the Chebychev
recursive relationship~Eq.~(\ref{cheb2}). Thus the cost of evaluating
$\bcal{V}_p$ is identical to the cost of carrying out the Ewald
sum. With the optimisation introduced here, this would imply a cost
that scales as $O({N^{1.5}})$. The issues of determining the number of
terms $N_\text{Ch}$, and the maximum and minimum eigenvalues of
$\bcal{D}$, must also be addressed before the total cost of Fixman's
procedure in the context of semidilute solutions can be estimated.

As pointed out earlier, it is not necessary to know the diffusion
matrix $\bcal{D}$ explicitly in order to describe the conformational
evolution of polymer molecules in a semidilute solution. However,
since \citet{beenakker} provides an expression for the periodic
diffusion tensor $\bm{\Tensor{D}}_{\nu \mu}$ in his original
derivation, it can used to determine the exact values of the maximum
and minimum eigenvalues, denoted here by
$\lambda_{\text{max}}^{\text{exact}}$ and
$\lambda_{\text{min}}^{\text{exact}}$. By comparing the values given
by Eqs.~(\ref{eq:lmax-lmin-dilute}) with the exact values (obtained
with the $\textsf{gsl}$\_$\textsf{eigen}$\_$\textsf{symm}$ subroutine
of the GNU Scientific Library), we find that the behaviour for our
semidilute system is quite different from what is known for
single-chain simulations: While in the latter case,
$\lambda_{\text{max}}^{\text{Fixman}}$ is a reasonable approximation
to $\lambda_{\text{max}}^{\text{exact}}$ (meaning that it scales in
the same way with the number of beads, with a constant ratio of order
unity), we here find that $\lambda_{\text{max}}^{\text{Fixman}}$ is
essentially independent of $N$, while
$\lambda_{\text{max}}^{\text{exact}}$ increases with $N$, roughly like
$N^{0.6}$. In other words, Eq.~(\ref{eq:lmax-lmin-dilute}) provides
only a poor approximation. The reason why the behaviour is so
different for dilute and semidilute systems is not clear to us; we
speculate that it might have to do with the different density
distributions of segments. Nevertheless, we can still use
$\lambda_{\text{max}}^{\text{Fixman}}$ for estimating the maximum
eigenvalue, since we empirically find, for a range of values of
$c/c^\star$, $a$, $N_b$ and $N_c$, and for a variety of polymer
conformations, the relation
\begin{equation}
\label{eq:fmax}
\lambda_{\text{max}}^{\text{exact}} = 0.35 \, N^{0.6} \,
\lambda_{\text{max}}^{\text{Fixman}}
\end{equation}
which is therefore used to estimate $\lambda_{\text{max}}$, assuming
that it is valid throughout. Similarly, we find empirically that the
lowest exact eigenvalue is essentially independent of the number of
segments, \ie
\begin{equation}
\label{eq:fmin}
\frac{\lambda_{\text{max}}^{\text{exact}}}
{\lambda_{\text{min}}^{\text{exact}}} 
= C \, N^{0.6}
\end{equation}
where the pre-factor $C$ depends on the values of $c/c^\star$, $a$ and
$N_b$, increasing with an increase in $a$ and $N_b$ and decreasing
with an increase in $c/c^\star$. For instance, for $c = 4.44 c^\star$,
$a = 0.5$ and $N_b = 10$, we find $C= 8$. It follows that in the
course of simulations a fairly accurate estimate of the minimum
eigenvalue can be obtained, once $\lambda_{\text{max}}$ is determined,
by using the expression $\lambda_{\text{min}} = \lambda_{\text{max}} /
(C \, N^{0.6})$. In general, the value of $C$ is obtained by trial and
error. Once $\lambda_{\text{max}}$ and $\lambda_{\text{min}}$ are
determined by this procedure, we find that it is adequate to use the
bounds $d_\textrm{max} = \lambda_{\text{max}} $ and $d_\textrm{min} =
\lambda_{\text{min}}$ to ensure a robust implementation of the
Chebychev polynomial approximation.

With regard to the number of Chebyshev terms, we find that for
semidilute solutions, as in the case of dilute solutions, the value of
$N_\text{Ch}$ required to keep the error in the estimation of the
square root below a fixed tolerance, scales as
$({\lambda_{\text{max}}}/{\lambda_{\text{min}} })^{\frac{1}{2}}$. This
immediately suggests from Eq.~(\ref{eq:fmin}) that the CPU-time
requirement for the calculation of $d \bcal{S}$ for semidilute
solutions using Fixman's method scales as $N_\text{Ch} N^{1.5} \sim
N^{1.8}$. Equation~(\ref{nint}) is used here to provide an initial
guess for $N_\text{Ch}$, which is then incrementally increased until
the relative error $E_f $ in the estimation of the square root
function, given by the following expression suggested in
Ref.~\onlinecite{jendrejack},
\begin{equation}
E_f = \left( \frac{\left\vert \, 
\left(\bcal{B} \cdot \Vector{\Delta} \bcal{W} \right) 
\cdot 
\left(\bcal{B} \cdot \Vector{\Delta} \bcal{W} \right) 
-
\Vector{\Delta} \bcal{W} \cdot \bcal{D} \cdot
\Vector{\Delta} \bcal{W}
\, \right\vert}
{\Vector{\Delta} \bcal{W} \cdot \bcal{D} \cdot
\Vector{\Delta} \bcal{W}
} \right)^{1/2}
\end{equation}
is less than a specified tolerance. In practice we find that the
choice of $C$ affects the efficiency with which the final value of
$N_\text{Ch}$ is obtained.

\section{\label{sec:euler-opt} Optimisation of Each
Euler Time Step }

The implementation of the Euler algorithm used here to determine the
configurational evolution of the system requires the estimation of the
drift term $\sum_{\mu} {\Tensor{D}}_{\nu \mu} \cdot {\Vector{F}}_{\mu}
$ and the diffusion term $\sum_\mu {\bm{\Tensor{B}}}_{\nu
  \mu} \cdot {\Vector{\Delta \bm{W}}}_{\mu}$ \emph{at each time step},
since the algorithm proceeds by evaluating the right hand side of
Eq.~(\ref{eq:sde}) for each bead $\nu$ in the original simulation
box. As mentioned earlier, determining the latter sum involves the
repeated invocation of the Ewald sum. Since the spatial configuration
of the system is frozen in a single time step, all terms in the Ewald
sum that are either (i) constant, (ii) only dependent on the
reciprocal space vector $\bf{k}$, or (iii) only dependent on the
spatial configuration, do not have to be repeatedly evaluated. As a
result, it becomes necessary not only to discuss the optimal
evaluation and scaling with system size of the drift and diffusion
terms individually (as we have in Secs.~\ref{sec:opt-DF} and
\ref{sec:chebyshev}), but also to consider the overall optimisation of
each Euler time step.

It turns out that there are two ways in which this optimisation can be
carried out. In order to give a flavour of the issues involved, we
only discuss here the treatment of the term involving $
\bm{\Tensor{M}}^{(1)}(\textbf{r}_{\nu\mu, \textbf{n}})$ in the Ewald
sum (see Eq.~(\ref{eq:DFmodified})). The remaining terms are either
treated similarly, or entail a more straightforward
treatment. Clearly, the term involving $
\bm{\Tensor{M}}^{(1)}(\textbf{r}_{\nu\mu, \textbf{n}})$ is first
evaluated when the drift term $\sum_{\mu} {\Tensor{D}}_{\nu \mu} \cdot
{\Vector{F}}_{\mu} $ is evaluated. Subsequently, it is required each
time the term $\bcal{D} \cdot {\bcal{V}}_p \,; \, p=0,\ldots,
N_\text{Ch}-1$ is evaluated in the Chebychev polynomial approximation
(see Eqs.~(\ref{cheb2}) and~(\ref{cheb3})). For ease of discussion, we
denote by ${\mathcal{V}}_{\mu}^{s}$ the $3N$ components of a typical
vector $ {\bcal{V}}_p$. Then the term involving $
\bm{\Tensor{M}}^{(1)}(\textbf{r}_{\nu\mu, \textbf{n}})$ in the
implementation of the Chebychev polynomial approximation can be
written as $ {\sum^{\prime}_{\textbf{n}}} \sum_{\mu = 1}^{N}
\sum_{s=1}^{3} M_{\nu \mu,\textbf{n}}^{rs}{\mathcal{V}}_{\mu}^{s}$,
where $M_{\nu \mu,\textbf{n}}^{rs}$ represents the
`$r\!s$'$^\text{th}$ Cartesian component of the tensor $
\bm{\Tensor{M}}^{(1)}(\textbf{r}_{\nu\mu, \textbf{n}})$. Before
discussing the two methods of optimisation used here, it is worth
noting the following point that is common to both methods. For each
bead $\nu$, in any periodic image ${\textbf{n}}$, the sum over the
index $\mu$ is carried out only over the nearest neighbours of bead
$\nu$, \ie, over the $N_{\text{r}_{\text{c}}}$ particles that lie
within a sphere centred at bead $\nu$ with cutoff radius $r_{\text{c}}$. The
choice of $r_{\text{c}}$, however, is different in the two schemes, as detailed
below.

In the first method of optimisation, which we denote here as HMA (for
``High Memory Algorithm''), the $3N \times 3N$ matrix $S_{\nu
  \mu}^{rs} = {\sum^{\prime}_{\textbf{n}}} M_{\nu
  \mu,\textbf{n}}^{rs}$ is calculated once and for all and stored in
the course of evaluating the drift term $\sum_{\mu} {\Tensor{D}}_{\nu
  \mu} \cdot {\Vector{F}}_{\mu} $. Note that the cost of evaluating
$S_{\nu \mu}^{rs}$ scales as $O(N \times N_{\text{r}_{\text{c}}})$
since in each periodic image ${\textbf{n}}$, only the beads $\mu$
whose distance from bead $\nu$ is less than a cutoff radius
${\left(r_{\text{c}}^{\text{HMA}}\right)}_{\text{opt}}$ are considered
in the sum over all periodic images. The nature of
${\left(r_{\text{c}}^{\text{HMA}}\right)}_{\text{opt}}$ and the value
of $N_{\text{r}_{\text{c}}}$ in this context, is discussed in more
detail below. It should be noted that the matrix $S_{\nu \mu}^{rs}$
becomes increasingly sparse when the system size is increased. While
it is therefore possible to save memory by sparse-matrix techniques
(meaning in practice the construction of a Verlet
table~\cite{vectorMD} and making use of indirect addressing), this was
not attempted here, \ie\ we stored the matrix with a simple $O(N^{2})$
implementation. Subsequently, each time the term $\bcal{D} \cdot
{\bcal{V}}_p \,; \, p=0,\ldots, N_\text{Ch}-1$ is calculated in the
Chebychev polynomial approximation, the $O(N^2)$ matrix multiplication
$ \sum_{\mu = 1}^{N} \sum_{s=1}^{3} S_{\nu
  \mu}^{rs}{\mathcal{V}}_{\mu}^{s}$ is carried out. Again, a sparse
matrix implementation might be able to reduce this computational
complexity. Ultimately this term dominates and the total CPU cost of
this scheme scales as $O(N_\text{Ch} \times N^2)$. For systems that
are not sufficiently large, the CPU cost might lie in the crossover
region between $O(N \times N_{\text{r}_{\text{c}}})$ (the cost for the
deterministic drift) and $O(N_\text{Ch} \times N^2)$.

\begin{figure*}[t]
\centering
\begin{tabular}{cc}
\includegraphics[width=0.48\textwidth]{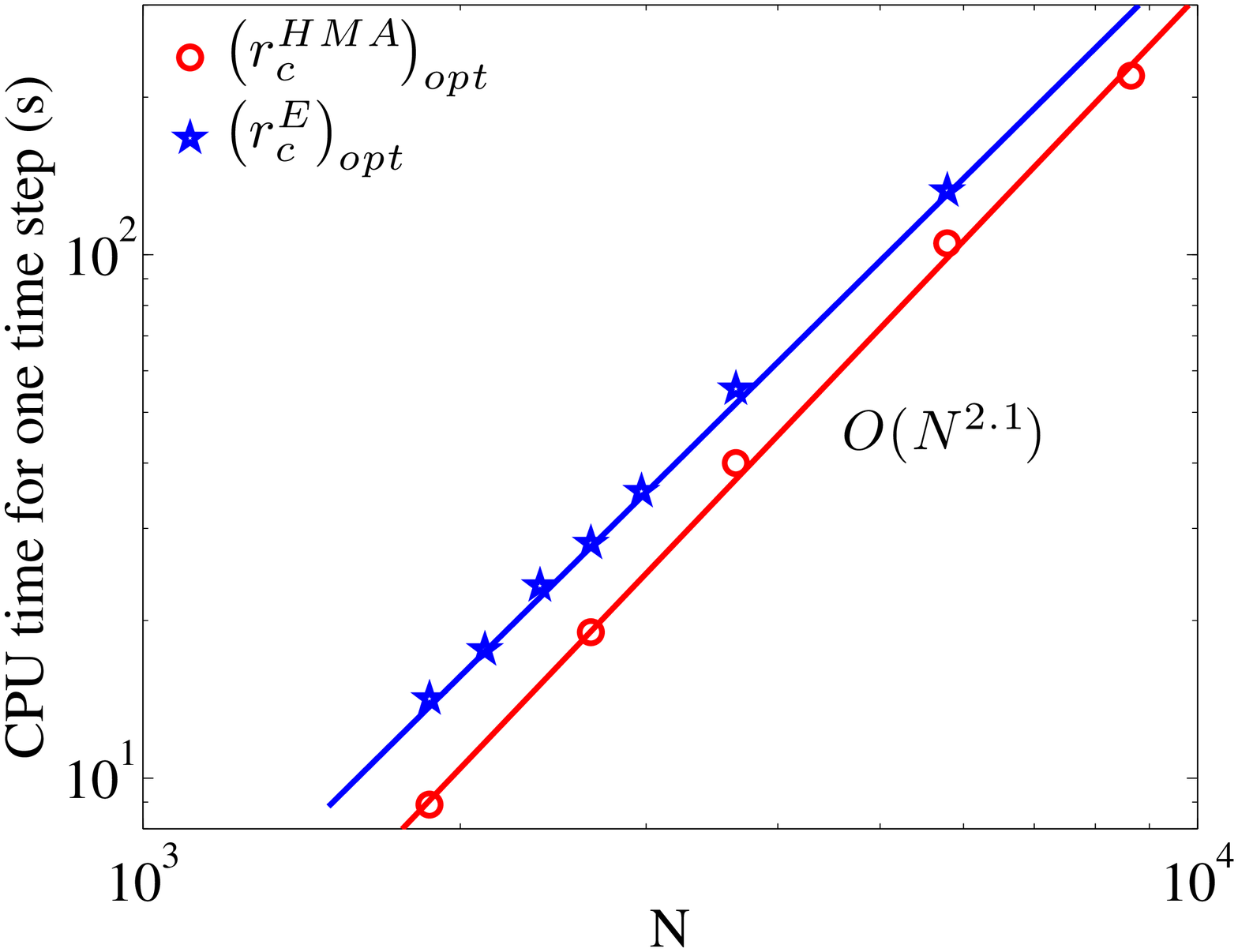} & 
\includegraphics[width=0.48\textwidth]{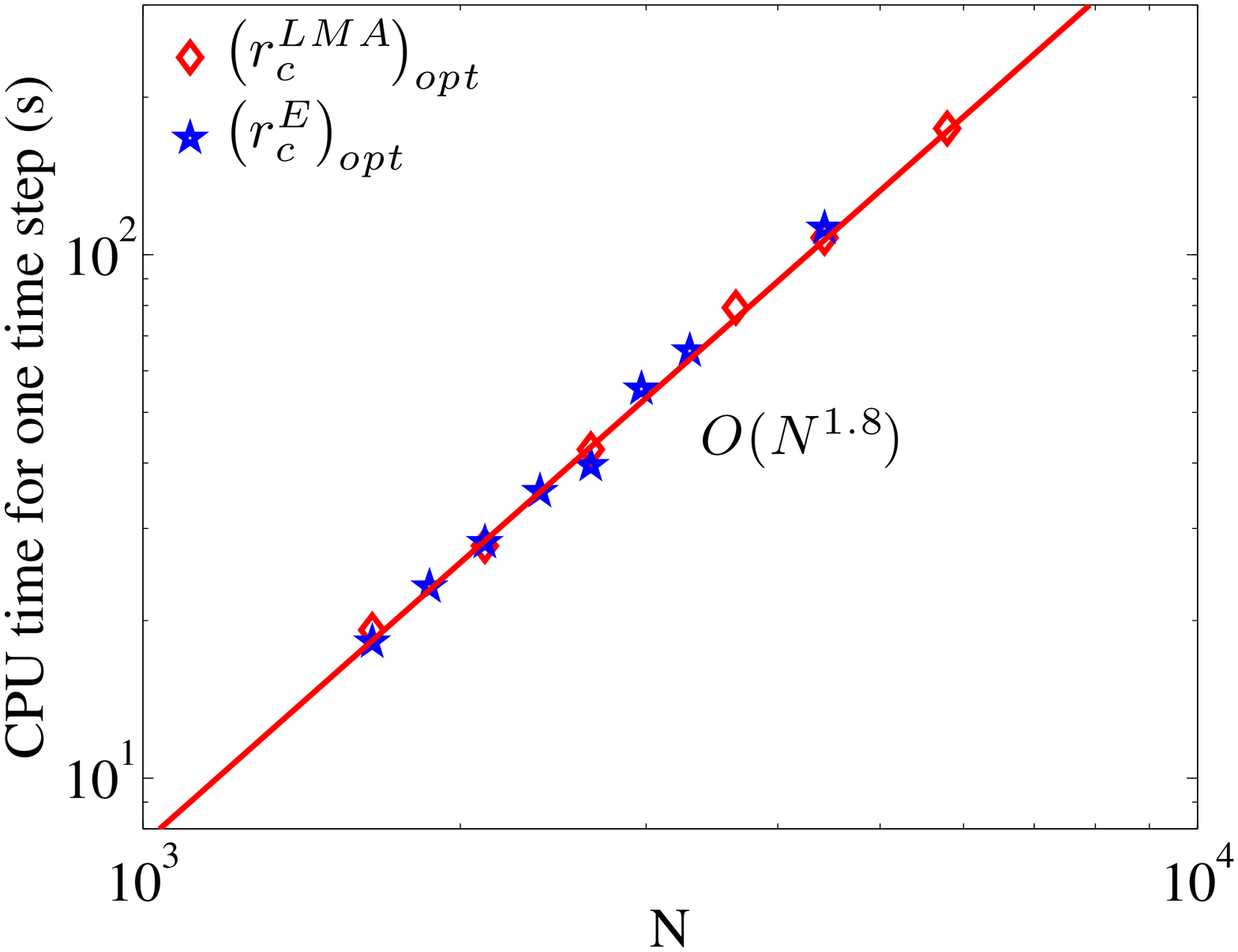} \\
(a) & (b)  \\
\end{tabular}
\caption{(Color online) \small \label{hmlm-scaling} CPU time scalings
  for (a) HMA (b) LMA. Symbols represent simulations results, and the
  lines are analytical estimates for the total time.}
\end{figure*}

The reason that
${\left(r_{\text{c}}^{\text{HMA}}\right)}_{\text{opt}}$ is different
from the cutoff radius
${\left(r_{\text{c}}^{\text{E}}\right)}_{\text{opt}}$ (calculated
earlier for just the evaluation of the Ewald sum) is because in the
HMA algorithm a different procedure is used in the repeated evaluation
of the Ewald sum, with certain quantities being calculated once and
for all and stored. By repeating the procedure adopted earlier for
optimising the bare Ewald sum, namely, by estimating the total time
required to evaluate the various quantities in the real space and
reciprocal space sums, we find for the CPU time for one step in
nanoseconds
\begin{eqnarray} 
\label{eq:THMA}
\nonumber
T^{\text{HMA}} / ns & = & 30 N(N-1) 
+ 1500 N \, r_{\text{c}}^3 c 
+ 23 N^2 (2 + N_{\text{Ch}})
\\
& + & 
0.67 \, \frac{M^6}{r_{\text{c}}^3 c} N^2 
\left(9.2 + 4.2 N_{\text{Ch}}\right) 
\end{eqnarray}   
where the constants reflect the various execution times for individual
terms on a 156 SGI Altix XE 320 cluster. Minimising this with respect
to the cutoff radius leads to
\begin{equation}
\label{eq:ropteuler1}
{\left(r_{\text{c}}^{\text{HMA}}\right)}_{\text{opt}}^3 c =
\frac{M^3}{8} \, {\left[0.25 N \, 
+ \, 0.12 N N_{\text{Ch}}\right]}^{1/2}
\end{equation} 
It turns out that ${\left(r_{\text{c}}^{\text{HMA}}\right)}_{\text{opt}} >
{\left(r_{\text{c}}^{\text{E}}\right)}_{\text{opt}}$. This is because a major
part of the real space calculation of ${\sum^{\prime}_{\textbf{n}}}
\, \bm{\Tensor{M}}^{(1)}(\textbf{r}_{\nu\mu, \textbf{n}})$ and
$\bm{\Tensor{M}}^{*}(\textbf{r}_{\nu\mu, \textbf{n = 0}})$ is not
repeated $N_{\text{Ch}}$ times in the HMA, leading to a cheaper real
space implementation. As a result, the optimisation procedure allows
the HMA to attribute a greater computational load to the real space
sum relative to the reciprocal space sum by having a larger cutoff
radius than ${\left(r_{\text{c}}^{\text{E}}\right)}_{\text{opt}}$. In contrast
to the bare Ewald sum, where $N_{\text{r}_{\text{c}}} \sim N^{0.5}$,
we find empirically that in the HMA, $N_{\text{r}_{\text{c}}} \sim
N^{0.7}$.

It is clear from Fig.~\ref{hmlm-scaling}~(a) that the CPU time for HMA
scales as $O(N^{2.1})$ when the simulation is run at a cutoff radius
of ${\left(r_{\text{c}}^{\text{HMA}}\right)}_{\text{opt}}$, with the
empirically estimated exponent 2.1 lying in the crossover regime
discussed earlier. Figure~\ref{hmlm-scaling}~(a) also indicates that
the CPU cost is greater when
${\left(r_{\text{c}}^{\text{E}}\right)}_{\text{opt}}$ is used in the HMA,
confirming the necessity to optimise the total procedure for
evaluating a single Euler time step rather than using the cutoff
radius obtained from the Ewald sum optimisation.

The major difference from the HMA, in the alternative method of
optimisation used here (denoted by LMA for ``Low Memory Algorithm''),
is the treatment of the sum $ {\sum^{\prime}_{\textbf{n}}} \sum_{\mu
  = 1}^{N} \sum_{s=1}^{3} M_{\nu
  \mu,\textbf{n}}^{rs}{\mathcal{V}}_{\mu}^{s}$. While many quantities,
such as those that are constant, or only functions of the reciprocal
space vector $\bf{k}$, are still calculated and stored once and for
all when the drift term $\sum_{\mu} {\Tensor{D}}_{\nu \mu} \cdot
{\Vector{F}}_{\mu} $ is evaluated, the $3N \times 3N$ matrix $S_{\nu
  \mu}^{rs}$ is not stored. Instead, the following two steps are
carried out: (i) For a given bead index $\nu$ and periodic image
${\textbf{n}}$, the quantity $T_{\nu,\textbf{n}}^{r} = \sum_{\mu =
  1}^{N} \sum_{s=1}^{3} M_{\nu
  \mu,\textbf{n}}^{rs}{\mathcal{V}}_{\mu}^{s}$ is evaluated, ensuring
that only those beads $\mu$ that lie within a cutoff radius
${\left(r_{\text{c}}^{\text{LMA}}\right)}_{\text{opt}}$ of bead $\nu$ are
considered in the sum over $\mu$. Note that for each bead $\mu$, the
sum over $s$ involves a simple $(3 \times 3) \times (3 \times 1)$
matrix multiplication. (ii) The sum $ {\sum^{\prime}_{\textbf{n}}}
\, T_{\nu,\textbf{n}}^{r}$ over periodic images ${\textbf{n}}$ is then
performed to obtain the required quantity in the Ewald sum.

Since, even in the LMA, some quantities are stored during the
evaluation of the drift term $\sum_{\mu} {\Tensor{D}}_{\nu \mu} \cdot
{\Vector{F}}_{\mu} $, we can optimise the entire process involved in
executing one time step in the Euler algorithm by choosing the
appropriate cutoff radius. Adopting the procedure described earlier,
we find for the CPU time per step in nanoseconds
\begin{eqnarray}
\label{eq:TLMA} 
\nonumber
T^{\text{LMA}} /ns & = & 1200 \, r_{\text{c}}^3 c  
\, N(1+N_{\text{Ch}})
\\
& + & 2 \, 
\frac{M^6}{r_{\text{c}}^3 c}  N^2\left(1.8 + 1.5 N_{\text{Ch}}\right) 
\end{eqnarray}
\begin{equation}
\label{eq:ropteuler}
{\left(r_{\text{c}}^{\text{LMA}}\right)}_{\text{opt}}^3 c
= \frac{M^3}{8} \, 
{\left[\frac{0.18 N \, + \, 0.15 N N_{\text{Ch}}}
{1 + N_{\text{Ch}}}\right]}^{1/2}
\end{equation} 
Figure~\ref{hmlm-scaling}~(b) compares the CPU cost involved when either
the cutoff radius ${\left(r_{\text{c}}^{\text{LMA}}\right)}_{\text{opt}}$ or
${\left(r_{\text{c}}^{\text{E}}\right)}_{\text{opt}}$ is used in the LMA. The
reason that ${\left(r_{\text{c}}^{\text{LMA}}\right)}_{\text{opt}}$ and
${\left(r_{\text{c}}^{\text{E}}\right)}_{\text{opt}}$ are nearly equal to each
other is because practically all the time consuming parts of the Ewald
sum are calculated repeatedly $N_{\text{Ch}}$ times in the LMA. As a
result, in contrast to the HMA, the saving achieved by storing some
quantities does not make a significant difference to the choice of
cutoff radius.
\begin{figure*}[t]
\centering
\begin{tabular}{cc}
\includegraphics[width=0.48\textwidth]{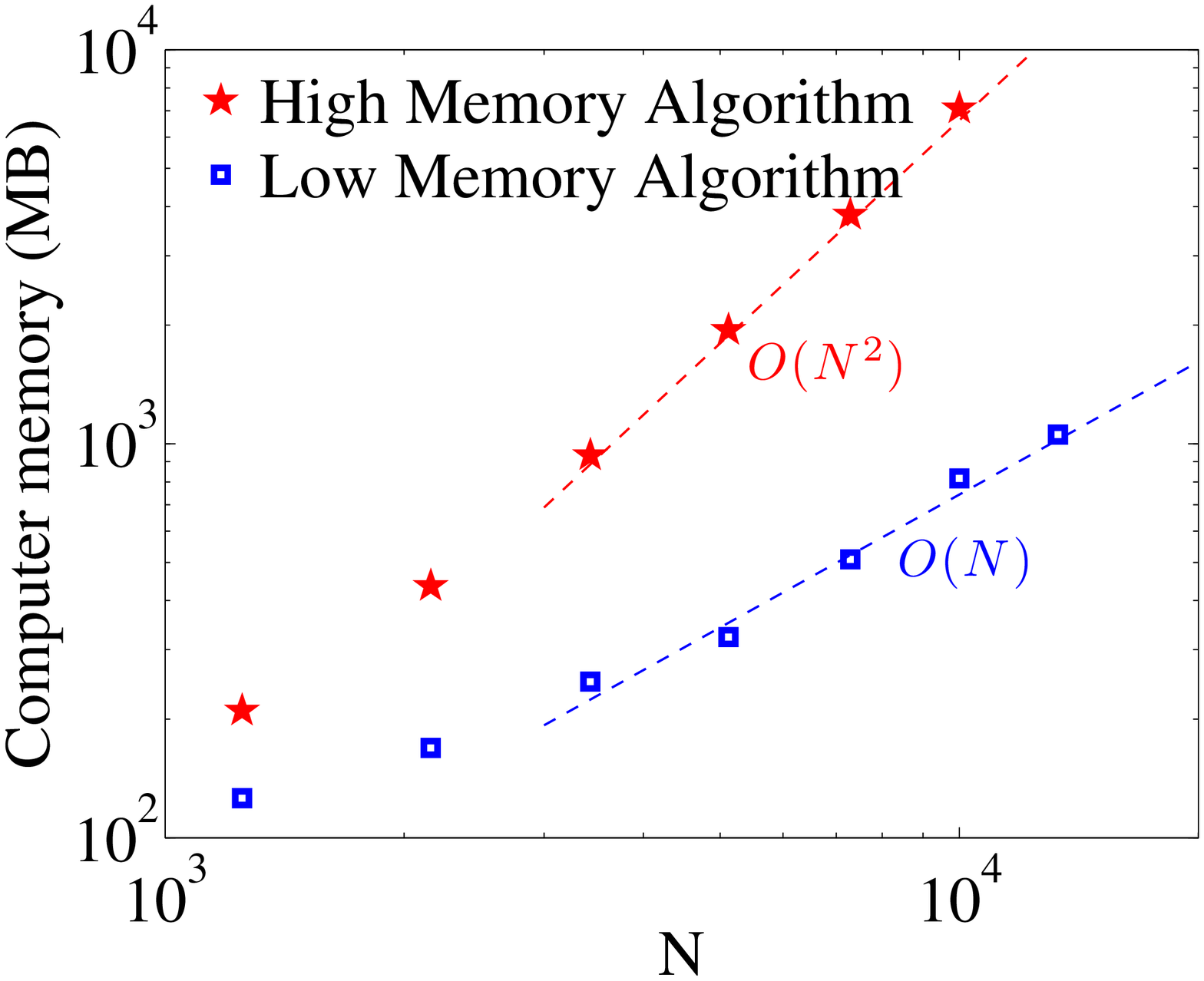} &
\includegraphics[width=0.48\textwidth]{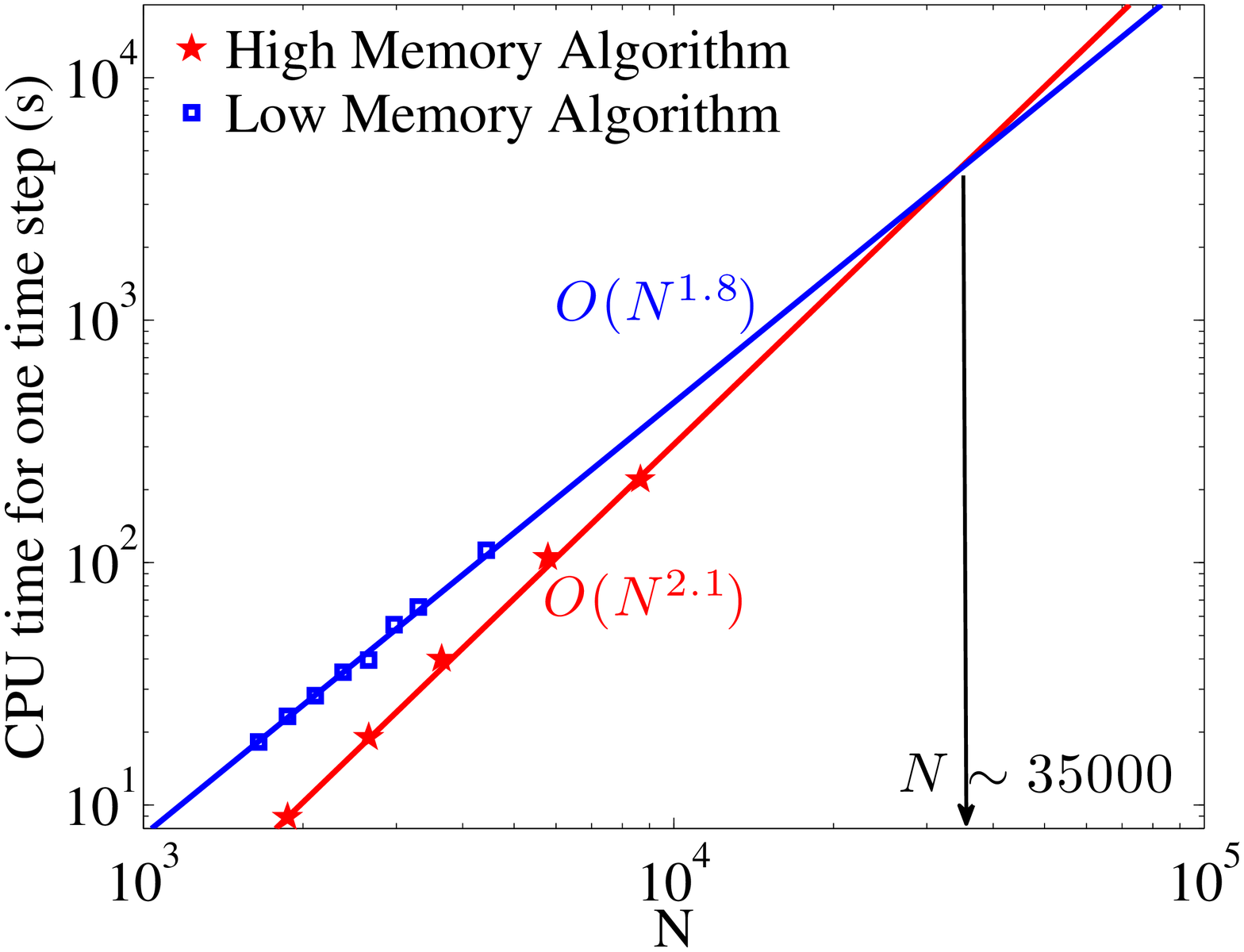} \\
(a) & (b)  \\
\end{tabular}
\caption{(Color online) \small \label{hmlm-comparison} Comparison
  between HMA and LMA for (a) Computer memory requirement (b) CPU time
  required for a single time step computation}
\end{figure*}

Unlike the HMA, there is no large storage requirement in the LMA, as
shown in Fig.~\ref{hmlm-comparison}~(a), where it is seen to scale as
$O(N)$ for sufficiently large $N$. Further, the CPU cost scales as $O(
N_\text{Ch} \times N \times N_{\text{r}_{\text{c}}})$, which is
identical to the scaling for the basic Ewald sum, namely, $O(N^{1.8})$
as can be seen in Fig.~\ref{hmlm-comparison}~(b). On the other hand,
since $S_{\nu \mu}^{rs}$ is not stored, the components $M_{\nu
  \mu,\textbf{n}}^{rs}$ are repeatedly evaluated in each of the
recursive Chebychev calculations. This extra calculation leads to a
large pre-factor in the scaling of the CPU with $N$. However, at $N
\sim 35000$ a crossover in the CPU cost can be seen to occur,
suggesting that it is advisable to use HMA below a system size of
roughly $35000$, while the LMA would be cheaper for larger systems.

%
\section{\label{sec:prop-validation} Testing
and Verification}

The optimised BD algorithm developed here is validated by testing and
verification under both $\theta$-solvent and good solvent
conditions. In the former case, we first check to see if static
equilibrium properties, namely, the radius of gyration and the
end-to-end vector, agree with known analytical results. Secondly, the
current implementation of the Ewald sum for hydrodynamic interactions
(which enables its use even in simulations that do not incorporate
excluded-volume interactions) is tested by comparing the prediction of
the infinite dilution equilibrium self-diffusion coefficient, which is
a dynamic property, with the results of a BD simulation of single
chain dynamics. As mentioned in Sec.~\ref{sec:intro}, we have
recently quantitatively compared the predictions of the explicit
solvent LB/MD method with the predictions of the implicit solvent BD
method for the dynamics of a single chain under good solvent
conditions in the dilute limit~\cite{tri}. A natural follow up of the
development of the current BD algorithm is to compare the two methods
at finite concentrations under good solvent conditions. Here we extend
our earlier study by comparing the predictions of the radius of
gyration, the end-to-end vector, and the self-diffusion
coefficient. This serves both to verify the predictions of the current
algorithm in a regime where there are no analytical predictions, and
to obtain an estimate of the relative computational costs of the two
mesoscopic simulation methods in the semidilute regime.

The mean-square end-to-end distance is given by
\begin{equation}
\langle R_e^2 \rangle = \langle {\left(\textbf{r}_{N_b}- 
\textbf{r}_1 \right)}^2 \rangle
\end{equation}
while the mean-square radius of gyration is given by
\begin{equation}
\langle R_g^2 \rangle = \frac{1}{2 N_b^2}
\sum_{\mu = 1}^{N_b} \sum_{\nu = 1}^{N_b} \langle  r_{\mu \nu}^2\rangle
\end{equation}
with $r_{\mu \nu} = \vert \textbf{r}_{\nu} - \textbf{r}_{\mu} \vert$
being the inter-particle distance. The long-time self-diffusion
coefficient is calculated by tracking the mean-square displacement of
the centre of mass $\textbf{r}_{c}$ of each chain
\begin{equation}
D_L = \lim_{t \to \infty} \left\langle \frac{{\vert\textbf{r}_{c} (t) 
- \textbf{r}_{c} (0)\vert}^2}{6 t} \right\rangle
\end{equation}

The predictions of the radius of gyration, the end-to-end vector, and
the self-diffusion coefficient by the current algorithm under
$\theta$-solvent and good solvent conditions, and their verification
by various means, are discussed in turn below.

\subsection{\label{sec:theta-validation} $\theta$-solvents}

\begin{figure*}[t]
\centering
\begin{tabular}{cc}
\includegraphics[width=0.48\textwidth]{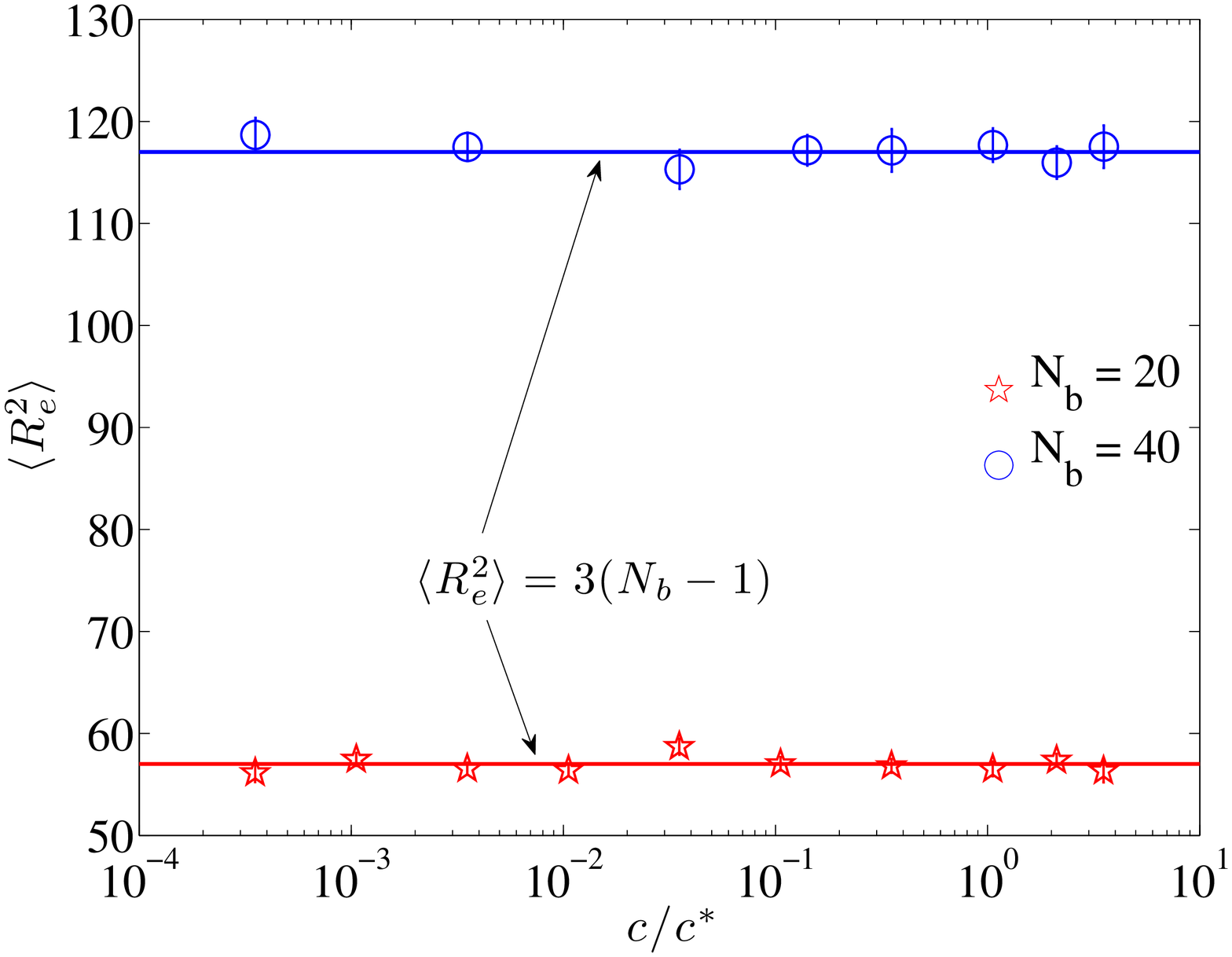} &
\includegraphics[width=0.48\textwidth]{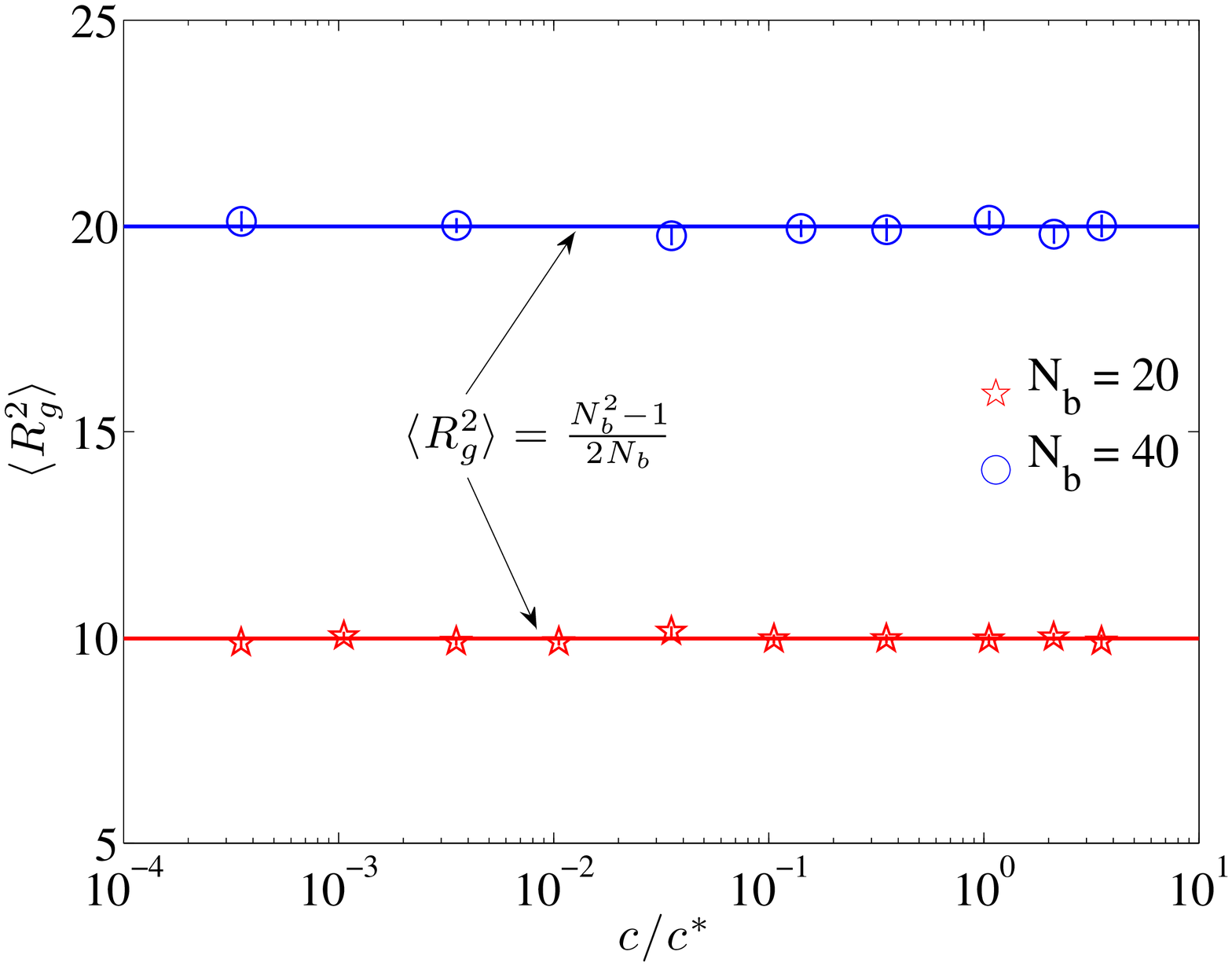} \\
(a) & (b)  \\
\end{tabular}
\caption{(Color online) \small \label{static-props} Validation of
  static properties under $\theta$-conditions: (a) Mean-square
  end-to-end distance $\langle R_e^2 \rangle$ (b) Mean-square radius
  of gyration $\langle R_g^2 \rangle$. Symbols indicate simulation
  data, while solid lines represent the analytical results given by
  Eqs.~(\ref{eq:Re2}) and (\ref{eq:Rg2}).}
\end{figure*}
The mean-square end-to-end distance and the mean-square radius of
gyration at equilibrium were obtained by carrying out simulations of
bead-spring chains with Hookean springs, using $N_b = 20$ and $40$ and
a fixed number of chains $N_c = 20$. The non-dimensional bead-radius
$a$ was chosen to be 0.5, and a time step $\Delta t = 0.01$ was used
to carry out the Euler integration. A range of concentrations from
$3 \times 10^{-4} \, c^\star$ to $3 \, c^\star$ were considered, with the
concentration being varied by changing the size of the simulation box
$L$. Since the chains are free to cross each other for
$\theta$-solvents, static properties such as the end-to-end distance
and the radius of gyration are independent of concentration. Further,
as is well known, their dependence on $N_{b}$ can be shown
analytically to be~\cite{bird}
\begin{equation}
\label{eq:Re2}
{\langle R_e^2 \rangle } = 3 (N_b - 1)
\end{equation}
and 
\begin{equation}
\label{eq:Rg2}
{\langle R_g^2 \rangle } = \frac{N_b^2 - 1}{2 N_b}
\end{equation}
Note that $c^\star$ can be determined once a choice for $N_{b}$ is made.
\begin{figure}[t]
\begin{center}
\includegraphics[width=0.48\textwidth]{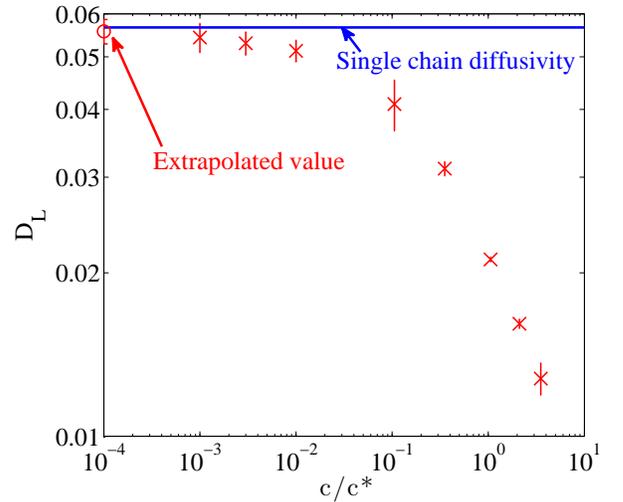}
\end{center}
\caption{(Color online) Long-time self-diffusion coefficient under
  $\theta$-solvent conditions. Symbols indicate simulation data
  obtained with the current multi-particle algorithm, while the solid
  line represents the value obtained by simulating the dynamics of a
  single chain in a dilute solution. The circle symbol on the $y$-axis
  is the value obtained by extrapolating the finite concentration
  results to the limit of zero concentration.}
\label{Dtheta}
\end{figure}

Figures~\ref{static-props}~(a) and (b) display the results for
$\langle R_e^2 \rangle$ and $\langle R_g^2 \rangle$, respectively, for
the range of concentrations considered here. Symbols indicate
simulation data, while solid lines represent the analytical results
given by Eqs.~(\ref{eq:Re2}) and (\ref{eq:Rg2}). Clearly, the
simulated static properties are in good agreement with analytical
predictions.
 
The single chain diffusion coefficient in a dilute solution under
$\theta$-solvent conditions is used here as the benchmark for
verifying the current implementation of the Ewald sum. The value of
the diffusion coefficient for $N_b = 20$ and $a = 0.5$ is displayed as
the solid line in Fig.~\ref{Dtheta}, obtained here by a conventional
BD simulation algorithm that uses a semi-implicit predictor corrector
scheme developed in our group for simulating a single chain that is
not confined in a box~\cite{prabhakar}. For the same set of input
simulation parameters, the long-time diffusivity is obtained from the
current multi-particle BD algorithm for a range of concentrations. It
is clear from Fig.~\ref{Dtheta} that the simulated data (symbols)
for the diffusivity $D_L$ approaches the single chain result in the
limit of zero concentration. The value of $D_L$ at $c/c^\star=0$ was
obtained by fitting the values at $c/c^\star = 0.001, 0.003$ and
$0.01$ with a second order polynomial and extrapolating to zero
concentration.

\begin{table*}[t]
  \caption{Comparison of predictions of the radius of gyration,
   the end-to-end vector, and
   the self-diffusion coefficient by the explicit solvent LB/MD
   method with the predictions
   of the implicit solvent BD method, for a bead-spring
   chain with $N_b = 10$ at three
   different concentrations, in a good solvent.
   Note that all properties are given in
   BD units, except the box size $L$, and concentration $c$,
   which are given in LB
   units when reported for the LB/MD simulations.
   Both $L$ and $c$ are identical
   in both methods when reported in the same
   unit system. Note that the highest concentration corresponds
   to melt-like conditions.
   \label{table1}}
 \begin{center}
{\small
\begin{tabular}{c c c c c c c c c}
\hline \hline

\parbox{5mm}{  \
\\ $N_c$ \\} 
&
\parbox{10mm}{  \
\\  Method \\} 
&
\parbox{15mm}{  \
\\ $L$ \\} 
&
\parbox{15mm}{  \
\\ $c$ \\} 
&
\parbox{22mm}{  \
\\ $c/c^\star$ \\} 
&
\parbox{22mm}{  \
\\ $\langle R_e^2 \rangle$ \\} 
&
\parbox{22mm}{  \
\\ $\langle R_g^2 \rangle$ \\} 
& 
\parbox{20mm}{  \
\\ $D_L$ \\} 
&
\\ \\
\hline \hline
20\,\,\,\,\,\,\,\,\,\,\,\,\vline  
& BD  & 24.152 & 0.0142 & 0.546$\pm$0.001 & 111.37$\pm$0.47 
& 18.36$\pm$0.04 & 0.0272$\pm$6$\times 10^{-4}$  \\
\cline{2-8}
\,\,\,\,\,\,\,\,\,\,\,\,\,\,\,\,\,\,\vline 
&  LB & 10 & 0.2 & 0.543$\pm$0.005 & 112.93$\pm$0.27	
& 18.29$\pm$0.02 & 0.0268$\pm$8$\times 10^{-4}$  \\
\hline \hline
32\,\,\,\,\,\,\,\,\,\,\,\,\vline 
&  BD & 21.737 & 0.0311 & 1.199$\pm$0.002 & 98.35$\pm$0.43 
& 16.6$\pm$0.04 & 0.0162$\pm$6$\times 10^{-4}$ \\
\cline{2-8}
\,\,\,\,\,\,\,\,\,\,\,\,\,\,\,\,\,\,\vline &  LB 
& 9 & 0.439 &  1.192$\pm$0.011 & 99.11$\pm$0.36 & 16.59$\pm$0.03 
& 0.0151$\pm$4$\times 10^{-4}$ \\
\hline \hline
70\,\,\,\,\,\,\,\,\,\,\,\,\vline 
&  BD & 21.737 & 0.068 & 2.623$\pm$0.004 & 76.07$\pm$0.53 
& 13.33$\pm$0.05 & 0.0024$\pm$1 $\times 10^{-4}$ \\
\cline{2-8}
\,\,\,\,\,\,\,\,\,\,\,\,\,\,\,\,\,\,\vline 
&  LB & 9 & 0.96 & 2.607$\pm$0.025 & 77.29$\pm$0.35 
& 13.42$\pm$0.04 & 0.00245$\pm$5$\times 10^{-5}$ \\
\hline \hline
\end{tabular}
}
\end{center}
\end{table*}

\subsection{\label{sec:good-validation} Good solvents}

In order to carry out a quantitative comparison between the LB/MD and
BD methods, it is necessary to ensure that the underlying polymer
model is identical for both the methods, and to map the input
parameters of the hybrid model onto the input values of the BD
model. A detailed discussion of how this can be achieved in the
context of dilute solutions has been given in
Ref.~\onlinecite{tri}. Exactly the same procedure has been adopted
here. Essentially, a bead-spring chain with FENE springs is used, with
a Weeks-Chandler-Andersen potential, which acts between all monomers,
employed to model the excluded-volume (EV) effect. While in the LB/MD
simulation approach, the Weeks-Chandler-Andersen parameters are used
to define the units of energy, length, and time, the corresponding
units in the BD simulations have been discussed earlier in
Sec.~\ref{sec:brownian}. We refer readers to Ref.~\onlinecite{tri}
for details of the length and time unit conversions between the two
methods. The comparison of the two methods proceeds by first picking
the simulation parameters for the LB/MD model, using these for the
LB/MD simulations, then converting them to BD units using the
procedure outlined in Ref.~\onlinecite{tri}, and finally running the
equivalent BD model obtained in this manner. In other words, the two
units systems are maintained in the respective methods, and a
comparison of predicted quantities carried out \emph{a posteriori}.

The results of carrying out this procedure for $\langle R_e^2
\rangle$, $\langle R_g^2 \rangle$ and $D_L$ are shown in
Table~(\ref{table1}) for $N_b = 10$ at three different
concentrations. It is worth noting that, since EV interactions are
short-ranged, we have implemented a neighbour-list in the current BD
algorithm for computing the pairwise summation of EV interactions,
with a cutoff radius equal to the range of the Weeks-Chandler-Andersen
potential. All values are reported in BD units, unless specified
otherwise. We find it convenient to maintain the same absolute
concentration in the two methods rather than the same $c/c^\star$, as this
would entail an interpolation procedure. In the BD method, $c^\star$ is
determined from $\langle R_g^2 \rangle$ in the dilute limit, by
carrying out a single chain simulation for parameter values that are
identical to those in the multi-particle BD simulation. In the LB/MD
method, simulations are carried out for three box sizes, $L=12$, $17$
and $21$, with the number of monomers held fixed (we use $N_{c} = 20$
and $N_{b}=10$). As a result, the monomer concentration decreases with
increasing box size. The values of $\langle R_g^2 \rangle$ obtained
for these three box sizes are extrapolated to infinite box size in
order to determine $\langle R_g^2 \rangle$ (and consequently $c^\star$) in
the dilute limit.

It is clear from Table~(\ref{table1}), both in the dilute limit, with
regard to values of $c/c^\star$, and at all three finite
concentrations, with regard to values of $\langle R_e^2 \rangle$,
$\langle R_g^2 \rangle$ and $D_L$, that there is excellent agreement
between the two mesoscopic simulation methods, since all properties
agree with each other within error bars. This validates the current
algorithm in a regime where there are no analytical
solutions. Further, it demonstrates the robustness of the parameter
mapping technique developed by \citet{tri} for comparing the two
simulation methods.

\section{\label{sec:CPUTIME-COMPARISON} 
Comparison of computational cost with LB/MD}
%
\begin{figure}[t]
\begin{center}
\includegraphics[width=0.48\textwidth]{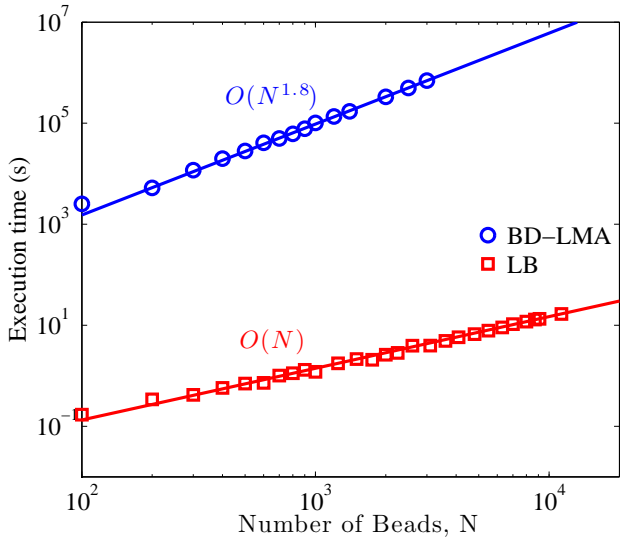}
\end{center}
\caption{(Color online) Comparison of the CPU time required by the LB
  and BD systems for a wide range of system sizes $N$, at
  concentration $c/c^\star = 1.2$, for the equivalent of one LB time
  unit.}
\label{BDLBCPUTIME}
\end{figure}

Our recent comparison~\cite{tri} of the predictions of the explicit
solvent LB/MD method with the predictions of the implicit solvent BD
method for the dynamics of a single chain indicated that in the dilute
limit, BD is the method of choice as it is significantly more
efficient than LB/MD. However, Fig.~\ref{BDLBCPUTIME} suggests that
for our current implementation the situation is quite the reverse at
the finite concentration, $c/c^\star=1.2$, at which the simulation
data in the figure were obtained. The comparison of the two mesoscopic
simulation methods displayed in Fig.~\ref{BDLBCPUTIME} was carried out
using the identical procedure developed earlier
by \citet{tri}. Essentially, the LB/MD method was run for a total of
100 MD time steps (with a step size of $0.01$ in LB units, or $0.018$
in BD units). This amounts to a total simulation time of one time unit
in terms of LB units. The BD algorithm was then run for the same
length of physical time, by converting one LB time unit to BD time
units. The BD algorithm required a significantly smaller time step of
$10^{-4}$ in BD units. The reason for this choice is because the
current implementation uses a simple Euler integration scheme, with a
rejection algorithm that ensures that none of the springs in any of
the bead-spring chains exceeds the upper limit of the FENE spring
length $\sqrt{b}$. In contrast, the earlier comparison of the two
methods in the dilute limit was based on a BD code that uses a
semi-implicit predictor-corrector method, enabling the use of a much
larger step size of $50 \times 10^{-4}$ BD units. The dependence of
CPU time on system size was examined here by increasing the number of
chains $N_{c}$, while keeping the number of beads in a chain fixed at
$N_{b} =10$. The concentration was maintained constant at
$c/c^\star=1.2$ (or $c = 0.031$ in BD units) by increasing the box
size $L$ suitably. Since the difference between the HMA and LMA BD
algorithms is insignificant on the scale of the difference between
LB/MD and BD, only results for the LMA are shown in
Fig.~\ref{BDLBCPUTIME}.

The CPU time scaling of the LMA algorithm has been established in
Sec.~\ref{sec:euler-opt} to be $N^{1.8}$. From Eqs.~(\ref{eq:TLMA})
and (\ref{eq:ropteuler}) one immediately sees that after optimisation
the CPU time depends only on the particle number $N$, but is
independent of the concentration $c$ (or the system volume $V$):
\begin{equation}
T^{\text{LMA}} (N, V) = \gamma^{\text{LMA}} N^{1.8}
\end{equation}
with some proportionality constant $\gamma^{\text{LMA}}$. Conversely, the
LB/MD method is dominated by the CPU effort of the solvent, \ie
\begin{equation}
T^{\text{LB}} (N, V) = \gamma^{\text{LB}} V = \gamma^{\text{LB}} \frac{N}{c}
\end{equation}
with another constant $\gamma^{\text{LB}}$. Hence
\begin{equation}
\frac{T^{\text{LMA}}}{T^{\text{LB}}} = 
\frac{\gamma^{\text{LMA}}}{\gamma^{\text{LB}}} c N^{0.8}
\end{equation}
From our CPU timings we find a value $\gamma^{\text{LMA}} /
\gamma^{\text{LB}} = 1.3 \times 10^{4}$ in BD units, \ie\ our
current implementation of Ewald BD becomes competitive with LB/MD only
if the concentration is below the very small value $7.8 \times 10^{-5}
\times N^{-0.8}$.

However, it should be noted that the present version is by far not the
fastest conceivable BD code. Firstly, we expect that by implementing
an implicit integrator the time step may be increased by nearly two
orders of magnitude. Secondly, the evaluation of the real-space HI
should be substantially faster (both in the LMA and HMA versions) by
making use of Verlet tables. Thirdly, the HMA algorithm could then
take advantage of sparse-matrix techniques (see also the discussion in
Sec.~\ref{sec:euler-opt}). Finally, the evaluation of the Fourier part
can be speeded up by making use of Fast Fourier Transformation, which,
as shown previously, gives rise to a complexity of the total algorithm
of $O(N^{1.3} \log N)$~\cite{sierou,banchio2003}. All together,
achieving accelerations by up to three orders of magnitude does not
seem unrealistic.

\section{\label{sec:summary} Summary}

A range of issues related to the development of an optimised BD
algorithm for simulating the dynamics of semidilute solutions in
unbounded domains has been considered here. In particular:
\begin{enumerate}
\item It is possible to develop an optimised Ewald method for
  hydrodynamic interactions that splits the cost of evaluating the
  real space and reciprocal space sums evenly, leading to a CPU cost
  that scales as $N^{1.5}$, rather than the $N^{2}$ scaling that would
  result from a straightforward implementation.
\item While Beenakker's original implementation of the Ewald sum is
  only valid for systems without bead overlap, it can be modified to
  account for bead overlap, such that $\theta$-solutions can be
  simulated by switching off excluded-volume interactions. To the best
  of our knowledge, this is the first implementation of an Ewald
  sum for the regularised branch of the RPY tensor.
\item As in the case of dilute solutions, the number of Chebychev
  terms required to maintain a desired accuracy scales as
  $({\lambda_{\text{max}}}/{\lambda_{\text{min}} })^{\frac{1}{2}}$,
  where $\lambda_{\text{max}}$ and $\lambda_{\text{min}}$ are the
  maximum and minimum eigenvalues of the diffusion tensor
  $\bcal{D}$. It is shown that this leads to an additional
  computational load that scales as $N^{0.3}$.
\item It is necessary to consider the optimisation of the overall time
  required to perform one Euler time step, in addition to the
  individual optimisations of the Ewald sum and Chebychev polynomial
  approximation. In this context, two different schemes for
  optimisation have been proposed in the form of the ``high memory''
  (HMA), and the ``low memory'' (LMA) algorithms. While the LMA leads
  to an overall CPU time scaling of $N^{1.8}$, which appears better
  than the $N^{2.1}$ scaling of the HMA, the large prefactor in the
  former makes it preferable only for large systems with more than
  roughly 35,000 particles.
\item The optimised BD algorithm gives accurate predictions under both
  $\theta$ and good solvent conditions. In the latter case, BD
  predictions are compared with those of the LB/MD method. The
  parameter mapping scheme developed by \citet{tri} for dilute
  solutions is found to be valid and useful even at a finite
  concentration in the semidilute regime.
\item In contrast to dilute solutions, where BD was shown to be
  significantly more computationally efficient than LB/MD~\cite{tri},
  exactly the opposite is true for semidilute solutions. The CPU cost
  of the BD method scales as $N^{1.8}$, while the cost of the LB/MD
  method scales linearly with system size. The necessity of carrying
  out an Ewald sum renders the BD method developed here significantly
  more computationally expensive than LB/MD. Nevertheless, it should
  be noted that the BD method can be further refined and dramatically
  speeded up, as discussed at the end of Sec.~\ref{sec:CPUTIME-COMPARISON}.
\end{enumerate}

\begin{acknowledgments} 
  The authors gratefully acknowledge CPU time grants from the National
  Computational Infrastructure (NCI) facility hosted by Australian
  National University, and Victorian Life Sciences Computation
  Initiative (VLSCI) hosted by University of Melbourne. The authors
  would also like to thank John Brady for very helpful discussions.
\end{acknowledgments}


\begin{thebibliography}{53}%
\makeatletter
\providecommand \@ifxundefined [1]{%
 \@ifx{#1\undefined}
}%
\providecommand \@ifnum [1]{%
 \ifnum #1\expandafter \@firstoftwo
 \else \expandafter \@secondoftwo
 \fi
}%
\providecommand \@ifx [1]{%
 \ifx #1\expandafter \@firstoftwo
 \else \expandafter \@secondoftwo
 \fi
}%
\providecommand \natexlab [1]{#1}%
\providecommand \enquote  [1]{``#1''}%
\providecommand \bibnamefont  [1]{#1}%
\providecommand \bibfnamefont [1]{#1}%
\providecommand \citenamefont [1]{#1}%
\providecommand \href@noop [0]{\@secondoftwo}%
\providecommand \href [0]{\begingroup \@sanitize@url \@href}%
\providecommand \@href[1]{\@@startlink{#1}\@@href}%
\providecommand \@@href[1]{\endgroup#1\@@endlink}%
\providecommand \@sanitize@url [0]{\catcode `\\12\catcode `\$12\catcode
  `\&12\catcode `\#12\catcode `\^12\catcode `\_12\catcode `\%12\relax}%
\providecommand \@@startlink[1]{}%
\providecommand \@@endlink[0]{}%
\providecommand \url  [0]{\begingroup\@sanitize@url \@url }%
\providecommand \@url [1]{\endgroup\@href {#1}{\urlprefix }}%
\providecommand \urlprefix  [0]{URL }%
\providecommand \Eprint [0]{\href }%
\providecommand \doibase [0]{http://dx.doi.org/}%
\providecommand \selectlanguage [0]{\@gobble}%
\providecommand \bibinfo  [0]{\@secondoftwo}%
\providecommand \bibfield  [0]{\@secondoftwo}%
\providecommand \translation [1]{[#1]}%
\providecommand \BibitemOpen [0]{}%
\providecommand \bibitemStop [0]{}%
\providecommand \bibitemNoStop [0]{.\EOS\space}%
\providecommand \EOS [0]{\spacefactor3000\relax}%
\providecommand \BibitemShut  [1]{\csname bibitem#1\endcsname}%
\let\auto@bib@innerbib\@empty
\bibitem [{\citenamefont
  {De~Gennes}(1976{\natexlab{a}})}]{de-gennes-macro1976-1}%
  \BibitemOpen
  \bibfield  {author} {\bibinfo {author} {\bibfnamefont {P.~G.}\ \bibnamefont
  {De~Gennes}},\ }\href {\doibase 10.1021/ma60052a011} {\bibfield  {journal}
  {\bibinfo  {journal} {Macromolecules}\ }\textbf {\bibinfo {volume} {9}},\
  \bibinfo {pages} {587} (\bibinfo {year} {1976}{\natexlab{a}})}\BibitemShut
  {NoStop}%
\bibitem [{\citenamefont {De~Gennes}(1976{\natexlab{b}})}]{de-gennes-paper}%
  \BibitemOpen
  \bibfield  {author} {\bibinfo {author} {\bibfnamefont {P.~G.}\ \bibnamefont
  {De~Gennes}},\ }\href {\doibase 10.1021/ma60052a012} {\bibfield  {journal}
  {\bibinfo  {journal} {Macromolecules}\ }\textbf {\bibinfo {volume} {9}},\
  \bibinfo {pages} {594} (\bibinfo {year} {1976}{\natexlab{b}})}\BibitemShut
  {NoStop}%
\bibitem [{\citenamefont {de~Gennes}(1979)}]{de-gennes}%
  \BibitemOpen
  \bibfield  {author} {\bibinfo {author} {\bibfnamefont {P.~G.}\ \bibnamefont
  {de~Gennes}},\ }\href@noop {} {\emph {\bibinfo {title} {Scaling Concepts in
  Polymer Physics}}}\ (\bibinfo  {publisher} {Cornell University},\ \bibinfo
  {address} {Ithaca, New York},\ \bibinfo {year} {1979})\BibitemShut {NoStop}%
\bibitem [{\citenamefont {Muthukumar}\ and\ \citenamefont
  {Edwards}(1982{\natexlab{a}})}]{muthukumar1982polymer}%
  \BibitemOpen
  \bibfield  {author} {\bibinfo {author} {\bibfnamefont {M.}~\bibnamefont
  {Muthukumar}}\ and\ \bibinfo {author} {\bibfnamefont {S.}~\bibnamefont
  {Edwards}},\ }\href {\doibase 10.1016/0032-3861(82)90333-0} {\bibfield
  {journal} {\bibinfo  {journal} {Polymer}\ }\textbf {\bibinfo {volume} {23}},\
  \bibinfo {pages} {345 } (\bibinfo {year} {1982}{\natexlab{a}})}\BibitemShut
  {NoStop}%
\bibitem [{\citenamefont {Muthukumar}\ and\ \citenamefont
  {Edwards}(1982{\natexlab{b}})}]{muthukumar1982JCP}%
  \BibitemOpen
  \bibfield  {author} {\bibinfo {author} {\bibfnamefont {M.}~\bibnamefont
  {Muthukumar}}\ and\ \bibinfo {author} {\bibfnamefont {S.~F.}\ \bibnamefont
  {Edwards}},\ }\href {\doibase 10.1063/1.443257} {\bibfield  {journal}
  {\bibinfo  {journal} {J. Chem. Phys.}\ }\textbf {\bibinfo {volume} {76}},\
  \bibinfo {pages} {2720} (\bibinfo {year} {1982}{\natexlab{b}})}\BibitemShut
  {NoStop}%
\bibitem [{\citenamefont {Muthukumar}\ and\ \citenamefont
  {Edwards}(1983)}]{muthukumar1983macro}%
  \BibitemOpen
  \bibfield  {author} {\bibinfo {author} {\bibfnamefont {M.}~\bibnamefont
  {Muthukumar}}\ and\ \bibinfo {author} {\bibfnamefont {S.~F.}\ \bibnamefont
  {Edwards}},\ }\href {\doibase 10.1021/ma00243a012} {\bibfield  {journal}
  {\bibinfo  {journal} {Macromolecules}\ }\textbf {\bibinfo {volume} {16}},\
  \bibinfo {pages} {1475} (\bibinfo {year} {1983})}\BibitemShut {NoStop}%
\bibitem [{\citenamefont {Edwards}\ and\ \citenamefont
  {Muthukumar}(1984)}]{muthukumar1984macro}%
  \BibitemOpen
  \bibfield  {author} {\bibinfo {author} {\bibfnamefont {S.~F.}\ \bibnamefont
  {Edwards}}\ and\ \bibinfo {author} {\bibfnamefont {M.}~\bibnamefont
  {Muthukumar}},\ }\href {\doibase 10.1021/ma00134a012} {\bibfield  {journal}
  {\bibinfo  {journal} {Macromolecules}\ }\textbf {\bibinfo {volume} {17}},\
  \bibinfo {pages} {586} (\bibinfo {year} {1984})}\BibitemShut {NoStop}%
\bibitem [{\citenamefont {Rubinstein}\ and\ \citenamefont
  {Colby}(2003)}]{rubinstein}%
  \BibitemOpen
  \bibfield  {author} {\bibinfo {author} {\bibfnamefont {M.}~\bibnamefont
  {Rubinstein}}\ and\ \bibinfo {author} {\bibfnamefont {R.~H.}\ \bibnamefont
  {Colby}},\ }\href@noop {} {\emph {\bibinfo {title} {Polymer Physics}}}\
  (\bibinfo  {publisher} {Oxford University Press},\ \bibinfo {year}
  {2003})\BibitemShut {NoStop}%
\bibitem [{\citenamefont {Ahlrichs}\ \emph {et~al.}(2001)\citenamefont
  {Ahlrichs}, \citenamefont {Everaers},\ and\ \citenamefont
  {D\"{u}nweg}}]{BD2001}%
  \BibitemOpen
  \bibfield  {author} {\bibinfo {author} {\bibfnamefont {P.}~\bibnamefont
  {Ahlrichs}}, \bibinfo {author} {\bibfnamefont {R.}~\bibnamefont {Everaers}},
  \ and\ \bibinfo {author} {\bibfnamefont {B.}~\bibnamefont {D\"{u}nweg}},\
  }\href {\doibase 10.1103/PhysRevE.64.040501} {\bibfield  {journal} {\bibinfo
  {journal} {Phys. Rev. E}\ }\textbf {\bibinfo {volume} {64}},\ \bibinfo
  {pages} {040501} (\bibinfo {year} {2001})}\BibitemShut {NoStop}%
\bibitem [{\citenamefont {Stoltz}\ \emph {et~al.}(2006)\citenamefont {Stoltz},
  \citenamefont {de~Pablo},\ and\ \citenamefont {Graham}}]{stoltz}%
  \BibitemOpen
  \bibfield  {author} {\bibinfo {author} {\bibfnamefont {C.}~\bibnamefont
  {Stoltz}}, \bibinfo {author} {\bibfnamefont {J.~J.}\ \bibnamefont
  {de~Pablo}}, \ and\ \bibinfo {author} {\bibfnamefont {M.~D.}\ \bibnamefont
  {Graham}},\ }\href {\doibase 10.1122/1.2167468} {\bibfield  {journal}
  {\bibinfo  {journal} {J. Rheol.}\ }\textbf {\bibinfo {volume} {50}},\
  \bibinfo {pages} {137} (\bibinfo {year} {2006})}\BibitemShut {NoStop}%
\bibitem [{\citenamefont {Huang}\ \emph {et~al.}(2010)\citenamefont {Huang},
  \citenamefont {Winkler}, \citenamefont {Sutmann},\ and\ \citenamefont
  {Gompper}}]{gompper2010}%
  \BibitemOpen
  \bibfield  {author} {\bibinfo {author} {\bibfnamefont {C.-C.}\ \bibnamefont
  {Huang}}, \bibinfo {author} {\bibfnamefont {R.~G.}\ \bibnamefont {Winkler}},
  \bibinfo {author} {\bibfnamefont {G.}~\bibnamefont {Sutmann}}, \ and\
  \bibinfo {author} {\bibfnamefont {G.}~\bibnamefont {Gompper}},\ }\href
  {\doibase 10.1021/ma101836x} {\bibfield  {journal} {\bibinfo  {journal}
  {Macromolecules}\ }\textbf {\bibinfo {volume} {43}},\ \bibinfo {pages}
  {10107} (\bibinfo {year} {2010})}\BibitemShut {NoStop}%
\bibitem [{\citenamefont {Kirkwood}\ and\ \citenamefont
  {Riseman}(1948)}]{kirkwood1948}%
  \BibitemOpen
  \bibfield  {author} {\bibinfo {author} {\bibfnamefont {J.~G.}\ \bibnamefont
  {Kirkwood}}\ and\ \bibinfo {author} {\bibfnamefont {J.}~\bibnamefont
  {Riseman}},\ }\href {\doibase 10.1063/1.1746947} {\bibfield  {journal}
  {\bibinfo  {journal} {J. Chem. Phys.}\ }\textbf {\bibinfo {volume} {16}},\
  \bibinfo {pages} {565} (\bibinfo {year} {1948})}\BibitemShut {NoStop}%
\bibitem [{\citenamefont {Freed}\ and\ \citenamefont
  {Edwards}(1974)}]{edwards-freedJCP1974}%
  \BibitemOpen
  \bibfield  {author} {\bibinfo {author} {\bibfnamefont {K.~F.}\ \bibnamefont
  {Freed}}\ and\ \bibinfo {author} {\bibfnamefont {S.~F.}\ \bibnamefont
  {Edwards}},\ }\href {\doibase DOI:10.1063/1.1682545} {\bibfield  {journal}
  {\bibinfo  {journal} {J. Chem. Phys.}\ }\textbf {\bibinfo {volume} {61}},\
  \bibinfo {pages} {3626} (\bibinfo {year} {1974})}\BibitemShut {NoStop}%
\bibitem [{\citenamefont {Bixon}(1976)}]{bixon1976}%
  \BibitemOpen
  \bibfield  {author} {\bibinfo {author} {\bibfnamefont {M.}~\bibnamefont
  {Bixon}},\ }\href {\doibase 10.1146/annurev.pc.27.100176.000433} {\bibfield
  {journal} {\bibinfo  {journal} {Annu. Rev. Phys. Chem.}\ }\textbf {\bibinfo
  {volume} {27}},\ \bibinfo {pages} {65} (\bibinfo {year} {1976})}\BibitemShut
  {NoStop}%
\bibitem [{\citenamefont {Ahlrichs}\ and\ \citenamefont
  {D\"{u}nweg}(1999)}]{BD1999}%
  \BibitemOpen
  \bibfield  {author} {\bibinfo {author} {\bibfnamefont {P.}~\bibnamefont
  {Ahlrichs}}\ and\ \bibinfo {author} {\bibfnamefont {B.}~\bibnamefont
  {D\"{u}nweg}},\ }\href {\doibase 10.1063/1.480156} {\bibfield  {journal}
  {\bibinfo  {journal} {J. Chem. Phys.}\ }\textbf {\bibinfo {volume} {111}},\
  \bibinfo {pages} {8225} (\bibinfo {year} {1999})}\BibitemShut {NoStop}%
\bibitem [{\citenamefont {D\"{u}nweg}\ and\ \citenamefont
  {Ladd}(2009)}]{BDTL2009}%
  \BibitemOpen
  \bibfield  {author} {\bibinfo {author} {\bibfnamefont {B.}~\bibnamefont
  {D\"{u}nweg}}\ and\ \bibinfo {author} {\bibfnamefont {A.~J.~C.}\ \bibnamefont
  {Ladd}},\ }\href@noop {} {\bibfield  {journal} {\bibinfo  {journal} {Adv.
  Polym. Sci.}\ }\textbf {\bibinfo {volume} {221}},\ \bibinfo {pages} {89}
  (\bibinfo {year} {2009})}\BibitemShut {NoStop}%
\bibitem [{\citenamefont {Malevanets}\ and\ \citenamefont
  {Kapral}(1999)}]{kapral1999JCP}%
  \BibitemOpen
  \bibfield  {author} {\bibinfo {author} {\bibfnamefont {A.}~\bibnamefont
  {Malevanets}}\ and\ \bibinfo {author} {\bibfnamefont {R.}~\bibnamefont
  {Kapral}},\ }\href {\doibase 10.1063/1.478857} {\bibfield  {journal}
  {\bibinfo  {journal} {J. Chem. Phys.}\ }\textbf {\bibinfo {volume} {110}},\
  \bibinfo {pages} {8605} (\bibinfo {year} {1999})}\BibitemShut {NoStop}%
\bibitem [{\citenamefont {Kapral}(2008)}]{kapral_bookchapter}%
  \BibitemOpen
  \bibfield  {author} {\bibinfo {author} {\bibfnamefont {R.}~\bibnamefont
  {Kapral}},\ }\enquote {\bibinfo {title} {Multiparticle collision dynamics:
  Simulation of complex systems on mesoscales},}\ in\ \href {\doibase
  10.1002/9780470371572.ch2} {\emph {\bibinfo {booktitle} {Advances in Chemical
  Physics}}}\ (\bibinfo {year} {2008})\ pp.\ \bibinfo {pages}
  {89--146}\BibitemShut {NoStop}%
\bibitem [{\citenamefont {Gompper}\ \emph {et~al.}(2009)\citenamefont
  {Gompper}, \citenamefont {Ihle}, \citenamefont {Kroll},\ and\ \citenamefont
  {Winkler}}]{gompper2009}%
  \BibitemOpen
  \bibfield  {author} {\bibinfo {author} {\bibfnamefont {G.}~\bibnamefont
  {Gompper}}, \bibinfo {author} {\bibfnamefont {T.}~\bibnamefont {Ihle}},
  \bibinfo {author} {\bibfnamefont {D.}~\bibnamefont {Kroll}}, \ and\ \bibinfo
  {author} {\bibfnamefont {R.}~\bibnamefont {Winkler}},\ }\href@noop {}
  {\bibfield  {journal} {\bibinfo  {journal} {Adv. Polym. Sci.}\ }\textbf
  {\bibinfo {volume} {221}},\ \bibinfo {pages} {1} (\bibinfo {year}
  {2009})}\BibitemShut {NoStop}%
\bibitem [{\citenamefont {\"{O}ttinger}(1996)}]{ottinger}%
  \BibitemOpen
  \bibfield  {author} {\bibinfo {author} {\bibfnamefont {H.~C.}\ \bibnamefont
  {\"{O}ttinger}},\ }\href@noop {} {\emph {\bibinfo {title} {Stochastic
  Processes in Polymeric Fluids}}}\ (\bibinfo  {publisher} {Springer},\
  \bibinfo {address} {Berlin},\ \bibinfo {year} {1996})\BibitemShut {NoStop}%
\bibitem [{\citenamefont {Pham}\ \emph {et~al.}(2009)\citenamefont {Pham},
  \citenamefont {Schiller}, \citenamefont {Prakash},\ and\ \citenamefont
  {D\"{u}nweg}}]{tri}%
  \BibitemOpen
  \bibfield  {author} {\bibinfo {author} {\bibfnamefont {T.~T.}\ \bibnamefont
  {Pham}}, \bibinfo {author} {\bibfnamefont {U.~D.}\ \bibnamefont {Schiller}},
  \bibinfo {author} {\bibfnamefont {J.~R.}\ \bibnamefont {Prakash}}, \ and\
  \bibinfo {author} {\bibfnamefont {B.}~\bibnamefont {D\"{u}nweg}},\ }\href
  {\doibase 10.1063/1.3251771} {\bibfield  {journal} {\bibinfo  {journal} {J.
  Chem. Phys.}\ }\textbf {\bibinfo {volume} {131}},\ \bibinfo {eid} {164114}
  (\bibinfo {year} {2009})}\BibitemShut {NoStop}%
\bibitem [{\citenamefont {Ladd}\ \emph {et~al.}(2009)\citenamefont {Ladd},
  \citenamefont {Kekre},\ and\ \citenamefont {Butler}}]{Butler2009PRE}%
  \BibitemOpen
  \bibfield  {author} {\bibinfo {author} {\bibfnamefont {A.~J.~C.}\
  \bibnamefont {Ladd}}, \bibinfo {author} {\bibfnamefont {R.}~\bibnamefont
  {Kekre}}, \ and\ \bibinfo {author} {\bibfnamefont {J.~E.}\ \bibnamefont
  {Butler}},\ }\href {\doibase 10.1103/PhysRevE.80.036704} {\bibfield
  {journal} {\bibinfo  {journal} {Phys. Rev. E}\ }\textbf {\bibinfo {volume}
  {80}},\ \bibinfo {pages} {036704} (\bibinfo {year} {2009})}\BibitemShut
  {NoStop}%
\bibitem [{\citenamefont {Doi}\ and\ \citenamefont
  {Edwards}(1986)}]{doi-edwards-book}%
  \BibitemOpen
  \bibfield  {author} {\bibinfo {author} {\bibfnamefont {M.}~\bibnamefont
  {Doi}}\ and\ \bibinfo {author} {\bibfnamefont {S.~F.}\ \bibnamefont
  {Edwards}},\ }\href@noop {} {\emph {\bibinfo {title} {The Theory of Polymer
  Dynamics}}}\ (\bibinfo  {publisher} {Clarendon Press},\ \bibinfo {address}
  {Oxford, New York},\ \bibinfo {year} {1986})\BibitemShut {NoStop}%
\bibitem [{\citenamefont {Fixman}(1986)}]{fixman}%
  \BibitemOpen
  \bibfield  {author} {\bibinfo {author} {\bibfnamefont {M.}~\bibnamefont
  {Fixman}},\ }\href {\doibase 10.1021/ma00158a043} {\bibfield  {journal}
  {\bibinfo  {journal} {Macromolecules}\ }\textbf {\bibinfo {volume} {19}},\
  \bibinfo {pages} {1204} (\bibinfo {year} {1986})}\BibitemShut {NoStop}%
\bibitem [{\citenamefont {Jendrejack}\ \emph {et~al.}(2000)\citenamefont
  {Jendrejack}, \citenamefont {Graham},\ and\ \citenamefont
  {de~Pablo}}]{jendrejack}%
  \BibitemOpen
  \bibfield  {author} {\bibinfo {author} {\bibfnamefont {R.~M.}\ \bibnamefont
  {Jendrejack}}, \bibinfo {author} {\bibfnamefont {M.~D.}\ \bibnamefont
  {Graham}}, \ and\ \bibinfo {author} {\bibfnamefont {J.~J.}\ \bibnamefont
  {de~Pablo}},\ }\href {\doibase 10.1063/1.1305884} {\bibfield  {journal}
  {\bibinfo  {journal} {J. Chem. Phys.}\ }\textbf {\bibinfo {volume} {113}},\
  \bibinfo {pages} {2894} (\bibinfo {year} {2000})}\BibitemShut {NoStop}%
\bibitem [{\citenamefont {Kr\"{o}ger}\ \emph {et~al.}(2000)\citenamefont
  {Kr\"{o}ger}, \citenamefont {Alba-Perez}, \citenamefont {Laso},\ and\
  \citenamefont {\"{O}ttinger}}]{kroger}%
  \BibitemOpen
  \bibfield  {author} {\bibinfo {author} {\bibfnamefont {M.}~\bibnamefont
  {Kr\"{o}ger}}, \bibinfo {author} {\bibfnamefont {A.}~\bibnamefont
  {Alba-Perez}}, \bibinfo {author} {\bibfnamefont {M.}~\bibnamefont {Laso}}, \
  and\ \bibinfo {author} {\bibfnamefont {H.~C.}\ \bibnamefont {\"{O}ttinger}},\
  }\href {\doibase 10.1063/1.1288803} {\bibfield  {journal} {\bibinfo
  {journal} {J. Chem. Phys.}\ }\textbf {\bibinfo {volume} {113}},\ \bibinfo
  {pages} {4767} (\bibinfo {year} {2000})}\BibitemShut {NoStop}%
\bibitem [{\citenamefont {Prabhakar}\ and\ \citenamefont
  {Prakash}(2004)}]{prabhakar}%
  \BibitemOpen
  \bibfield  {author} {\bibinfo {author} {\bibfnamefont {R.}~\bibnamefont
  {Prabhakar}}\ and\ \bibinfo {author} {\bibfnamefont {J.}~\bibnamefont
  {Prakash}},\ }\href {\doibase 10.1016/S0377-0257(03)00155-1} {\bibfield
  {journal} {\bibinfo  {journal} {J. Non-Newtonian Fluid Mech.}\ }\textbf
  {\bibinfo {volume} {116}},\ \bibinfo {pages} {163 } (\bibinfo {year}
  {2004})}\BibitemShut {NoStop}%
\bibitem [{\citenamefont {Hasimoto}(1959)}]{hasimoto}%
  \BibitemOpen
  \bibfield  {author} {\bibinfo {author} {\bibfnamefont {H.}~\bibnamefont
  {Hasimoto}},\ }\href {\doibase 10.1017/S0022112059000222} {\bibfield
  {journal} {\bibinfo  {journal} {J. Fluid Mech.}\ }\textbf {\bibinfo {volume}
  {5}},\ \bibinfo {pages} {317} (\bibinfo {year} {1959})}\BibitemShut {NoStop}%
\bibitem [{\citenamefont {Beenakker}(1986)}]{beenakker}%
  \BibitemOpen
  \bibfield  {author} {\bibinfo {author} {\bibfnamefont {C.~W.~J.}\
  \bibnamefont {Beenakker}},\ }\href {\doibase 10.1063/1.451199} {\bibfield
  {journal} {\bibinfo  {journal} {J. Chem. Phys.}\ }\textbf {\bibinfo {volume}
  {85}},\ \bibinfo {pages} {1581} (\bibinfo {year} {1986})}\BibitemShut
  {NoStop}%
\bibitem [{\citenamefont {Ewald}(1921)}]{ewald}%
  \BibitemOpen
  \bibfield  {author} {\bibinfo {author} {\bibfnamefont {P.~P.}\ \bibnamefont
  {Ewald}},\ }\href {\doibase 10.1002/andp.19213690304} {\bibfield  {journal}
  {\bibinfo  {journal} {Annalen der Physik}\ }\textbf {\bibinfo {volume}
  {369}},\ \bibinfo {pages} {253} (\bibinfo {year} {1921})}\BibitemShut
  {NoStop}%
\bibitem [{\citenamefont {Allen}\ and\ \citenamefont
  {Tildesley}(1990)}]{allen}%
  \BibitemOpen
  \bibfield  {author} {\bibinfo {author} {\bibfnamefont {M.}~\bibnamefont
  {Allen}}\ and\ \bibinfo {author} {\bibfnamefont {D.}~\bibnamefont
  {Tildesley}},\ }\href@noop {} {\emph {\bibinfo {title} {Computer Simulations
  of Liquids}}}\ (\bibinfo  {publisher} {Oxford Science},\ \bibinfo {address}
  {London},\ \bibinfo {year} {1990})\BibitemShut {NoStop}%
\bibitem [{\citenamefont {Luty}\ \emph {et~al.}(1994)\citenamefont {Luty},
  \citenamefont {Davis}, \citenamefont {Tironi},\ and\ \citenamefont
  {Van~Gunsteren}}]{luty}%
  \BibitemOpen
  \bibfield  {author} {\bibinfo {author} {\bibfnamefont {B.~A.}\ \bibnamefont
  {Luty}}, \bibinfo {author} {\bibfnamefont {M.~E.}\ \bibnamefont {Davis}},
  \bibinfo {author} {\bibfnamefont {I.~G.}\ \bibnamefont {Tironi}}, \ and\
  \bibinfo {author} {\bibfnamefont {W.~F.}\ \bibnamefont {Van~Gunsteren}},\
  }\href {\doibase 10.1080/08927029408022004} {\bibfield  {journal} {\bibinfo
  {journal} {Mol. Simul.}\ }\textbf {\bibinfo {volume} {14}},\ \bibinfo {pages}
  {11} (\bibinfo {year} {1994})}\BibitemShut {NoStop}%
\bibitem [{\citenamefont {Toukmaji}\ and\ \citenamefont {{Board
  Jr}.}(1996)}]{toukmaji}%
  \BibitemOpen
  \bibfield  {author} {\bibinfo {author} {\bibfnamefont {A.~Y.}\ \bibnamefont
  {Toukmaji}}\ and\ \bibinfo {author} {\bibfnamefont {J.~A.}\ \bibnamefont
  {{Board Jr}.}},\ }\href {\doibase 10.1016/0010-4655(96)00016-1} {\bibfield
  {journal} {\bibinfo  {journal} {Comput. Phys. Commun.}\ }\textbf {\bibinfo
  {volume} {95}},\ \bibinfo {pages} {73 } (\bibinfo {year} {1996})}\BibitemShut
  {NoStop}%
\bibitem [{\citenamefont {Deserno}\ and\ \citenamefont {Holm}(1998)}]{holm}%
  \BibitemOpen
  \bibfield  {author} {\bibinfo {author} {\bibfnamefont {M.}~\bibnamefont
  {Deserno}}\ and\ \bibinfo {author} {\bibfnamefont {C.}~\bibnamefont {Holm}},\
  }\href {\doibase 10.1063/1.477414} {\bibfield  {journal} {\bibinfo  {journal}
  {J. Chem. Phys.}\ }\textbf {\bibinfo {volume} {109}},\ \bibinfo {pages}
  {7678} (\bibinfo {year} {1998})}\BibitemShut {NoStop}%
\bibitem [{\citenamefont {Smith}\ \emph {et~al.}(1987)\citenamefont {Smith},
  \citenamefont {Snook},\ and\ \citenamefont {Megen}}]{snook1987}%
  \BibitemOpen
  \bibfield  {author} {\bibinfo {author} {\bibfnamefont {E.}~\bibnamefont
  {Smith}}, \bibinfo {author} {\bibfnamefont {I.}~\bibnamefont {Snook}}, \ and\
  \bibinfo {author} {\bibfnamefont {W.~V.}\ \bibnamefont {Megen}},\ }\href
  {\doibase 10.1016/0378-4371(87)90160-9} {\bibfield  {journal} {\bibinfo
  {journal} {Phys. A: Stat. Mech. Appl.}\ }\textbf {\bibinfo {volume} {143}},\
  \bibinfo {pages} {441 } (\bibinfo {year} {1987})}\BibitemShut {NoStop}%
\bibitem [{\citenamefont {Brady}\ \emph {et~al.}(1988)\citenamefont {Brady},
  \citenamefont {Phillips}, \citenamefont {Lester},\ and\ \citenamefont
  {Bossis}}]{bplb}%
  \BibitemOpen
  \bibfield  {author} {\bibinfo {author} {\bibfnamefont {J.~F.}\ \bibnamefont
  {Brady}}, \bibinfo {author} {\bibfnamefont {R.~J.}\ \bibnamefont {Phillips}},
  \bibinfo {author} {\bibfnamefont {J.~C.}\ \bibnamefont {Lester}}, \ and\
  \bibinfo {author} {\bibfnamefont {G.}~\bibnamefont {Bossis}},\ }\href
  {\doibase 10.1017/S0022112088002411} {\bibfield  {journal} {\bibinfo
  {journal} {J. Fluid Mech.}\ }\textbf {\bibinfo {volume} {195}},\ \bibinfo
  {pages} {257} (\bibinfo {year} {1988})}\BibitemShut {NoStop}%
\bibitem [{\citenamefont {Rinn}\ \emph {et~al.}(1999)\citenamefont {Rinn},
  \citenamefont {Zahn}, \citenamefont {Maass},\ and\ \citenamefont
  {Maret}}]{maret1999}%
  \BibitemOpen
  \bibfield  {author} {\bibinfo {author} {\bibfnamefont {B.}~\bibnamefont
  {Rinn}}, \bibinfo {author} {\bibfnamefont {K.}~\bibnamefont {Zahn}}, \bibinfo
  {author} {\bibfnamefont {P.}~\bibnamefont {Maass}}, \ and\ \bibinfo {author}
  {\bibfnamefont {G.}~\bibnamefont {Maret}},\ }\href
  {http://stacks.iop.org/0295-5075/46/i=4/a=537} {\bibfield  {journal}
  {\bibinfo  {journal} {Europhys. Lett.}\ }\textbf {\bibinfo {volume} {46}},\
  \bibinfo {pages} {537} (\bibinfo {year} {1999})}\BibitemShut {NoStop}%
\bibitem [{\citenamefont {Sierou}\ and\ \citenamefont {Brady}(2001)}]{sierou}%
  \BibitemOpen
  \bibfield  {author} {\bibinfo {author} {\bibfnamefont {A.}~\bibnamefont
  {Sierou}}\ and\ \bibinfo {author} {\bibfnamefont {J.~F.}\ \bibnamefont
  {Brady}},\ }\href {\doibase 10.1017/S0022112001005912} {\bibfield  {journal}
  {\bibinfo  {journal} {J. Fluid Mech.}\ }\textbf {\bibinfo {volume} {448}},\
  \bibinfo {pages} {115} (\bibinfo {year} {2001})}\BibitemShut {NoStop}%
\bibitem [{\citenamefont {Banchio}\ and\ \citenamefont
  {Brady}(2003)}]{banchio2003}%
  \BibitemOpen
  \bibfield  {author} {\bibinfo {author} {\bibfnamefont {A.~J.}\ \bibnamefont
  {Banchio}}\ and\ \bibinfo {author} {\bibfnamefont {J.~F.}\ \bibnamefont
  {Brady}},\ }\href {\doibase 10.1063/1.1571819} {\bibfield  {journal}
  {\bibinfo  {journal} {J. Chem. Phys.}\ }\textbf {\bibinfo {volume} {118}},\
  \bibinfo {pages} {10323} (\bibinfo {year} {2003})}\BibitemShut {NoStop}%
\bibitem [{\citenamefont {Hern\'andez-Ortiz}\ \emph {et~al.}(2007)\citenamefont
  {Hern\'andez-Ortiz}, \citenamefont {de~Pablo},\ and\ \citenamefont
  {Graham}}]{ggem-graham2012}%
  \BibitemOpen
  \bibfield  {author} {\bibinfo {author} {\bibfnamefont {J.~P.}\ \bibnamefont
  {Hern\'andez-Ortiz}}, \bibinfo {author} {\bibfnamefont {J.~J.}\ \bibnamefont
  {de~Pablo}}, \ and\ \bibinfo {author} {\bibfnamefont {M.~D.}\ \bibnamefont
  {Graham}},\ }\href {\doibase 10.1103/PhysRevLett.98.140602} {\bibfield
  {journal} {\bibinfo  {journal} {Phys. Rev. Lett.}\ }\textbf {\bibinfo
  {volume} {98}},\ \bibinfo {pages} {140602} (\bibinfo {year}
  {2007})}\BibitemShut {NoStop}%
\bibitem [{\citenamefont {Perram}\ \emph {et~al.}(1988)\citenamefont {Perram},
  \citenamefont {Petersen},\ and\ \citenamefont {De~Leeuw}}]{perram}%
  \BibitemOpen
  \bibfield  {author} {\bibinfo {author} {\bibfnamefont {J.~W.}\ \bibnamefont
  {Perram}}, \bibinfo {author} {\bibfnamefont {H.~G.}\ \bibnamefont
  {Petersen}}, \ and\ \bibinfo {author} {\bibfnamefont {S.~W.}\ \bibnamefont
  {De~Leeuw}},\ }\href {\doibase 10.1080/00268978800101471} {\bibfield
  {journal} {\bibinfo  {journal} {Mol. Phys.}\ }\textbf {\bibinfo {volume}
  {65}},\ \bibinfo {pages} {875} (\bibinfo {year} {1988})}\BibitemShut
  {NoStop}%
\bibitem [{\citenamefont {Fincham}(1994)}]{fincham}%
  \BibitemOpen
  \bibfield  {author} {\bibinfo {author} {\bibfnamefont {D.}~\bibnamefont
  {Fincham}},\ }\href {\doibase 10.1080/08927029408022180} {\bibfield
  {journal} {\bibinfo  {journal} {Mol. Simul.}\ }\textbf {\bibinfo {volume}
  {13}},\ \bibinfo {pages} {1} (\bibinfo {year} {1994})}\BibitemShut {NoStop}%
\bibitem [{\citenamefont {Rotne}\ and\ \citenamefont {Prager}(1969)}]{rpy}%
  \BibitemOpen
  \bibfield  {author} {\bibinfo {author} {\bibfnamefont {J.}~\bibnamefont
  {Rotne}}\ and\ \bibinfo {author} {\bibfnamefont {S.}~\bibnamefont {Prager}},\
  }\href {\doibase 10.1063/1.1670977} {\bibfield  {journal} {\bibinfo
  {journal} {J. Chem. Phys.}\ }\textbf {\bibinfo {volume} {50}},\ \bibinfo
  {pages} {4831} (\bibinfo {year} {1969})}\BibitemShut {NoStop}%
\bibitem [{\citenamefont {Yamakawa}(1970)}]{yamakawa70}%
  \BibitemOpen
  \bibfield  {author} {\bibinfo {author} {\bibfnamefont {H.}~\bibnamefont
  {Yamakawa}},\ }\href@noop {} {\bibfield  {journal} {\bibinfo  {journal} {J.
  Chem. Phys.}\ }\textbf {\bibinfo {volume} {53}},\ \bibinfo {pages} {436}
  (\bibinfo {year} {1970})}\BibitemShut {NoStop}%
\bibitem [{\citenamefont {Mazur}\ and\ \citenamefont {van
  Saarloos}(1982)}]{MazurVanSaarloos}%
  \BibitemOpen
  \bibfield  {author} {\bibinfo {author} {\bibfnamefont {P.}~\bibnamefont
  {Mazur}}\ and\ \bibinfo {author} {\bibfnamefont {W.}~\bibnamefont {van
  Saarloos}},\ }\href {\doibase 10.1016/0378-4371(82)90127-3} {\bibfield
  {journal} {\bibinfo  {journal} {Physica A}\ }\textbf {\bibinfo {volume}
  {115}},\ \bibinfo {pages} {21 } (\bibinfo {year} {1982})}\BibitemShut
  {NoStop}%
\bibitem [{\citenamefont {Sunthar}\ and\ \citenamefont
  {Prakash}(2005)}]{suntharmacro}%
  \BibitemOpen
  \bibfield  {author} {\bibinfo {author} {\bibfnamefont {P.}~\bibnamefont
  {Sunthar}}\ and\ \bibinfo {author} {\bibfnamefont {J.~R.}\ \bibnamefont
  {Prakash}},\ }\href {\doibase 10.1021/ma035941l} {\bibfield  {journal}
  {\bibinfo  {journal} {Macromolecules}\ }\textbf {\bibinfo {volume} {38}},\
  \bibinfo {pages} {617} (\bibinfo {year} {2005})}\BibitemShut {NoStop}%
\bibitem [{\citenamefont {Zylka}\ and\ \citenamefont
  {Ottinger}(1989)}]{zylka89}%
  \BibitemOpen
  \bibfield  {author} {\bibinfo {author} {\bibfnamefont {W.}~\bibnamefont
  {Zylka}}\ and\ \bibinfo {author} {\bibfnamefont {H.~C.}\ \bibnamefont
  {Ottinger}},\ }\href {\doibase 10.1063/1.456690} {\bibfield  {journal}
  {\bibinfo  {journal} {J. Chem. Phys.}\ }\textbf {\bibinfo {volume} {90}},\
  \bibinfo {pages} {474} (\bibinfo {year} {1989})}\BibitemShut {NoStop}%
\bibitem [{\citenamefont {Thurston}\ and\ \citenamefont
  {Peterlin}(1967)}]{thurston1967}%
  \BibitemOpen
  \bibfield  {author} {\bibinfo {author} {\bibfnamefont {G.~B.}\ \bibnamefont
  {Thurston}}\ and\ \bibinfo {author} {\bibfnamefont {A.}~\bibnamefont
  {Peterlin}},\ }\href {\doibase 10.1063/1.1840651} {\bibfield  {journal}
  {\bibinfo  {journal} {J. Chem. Phys.}\ }\textbf {\bibinfo {volume} {46}},\
  \bibinfo {pages} {4881} (\bibinfo {year} {1967})}\BibitemShut {NoStop}%
\bibitem [{\citenamefont {Bird}\ \emph {et~al.}(1987)\citenamefont {Bird},
  \citenamefont {Curtiss}, \citenamefont {Armstrong},\ and\ \citenamefont
  {Hassager}}]{bird}%
  \BibitemOpen
  \bibfield  {author} {\bibinfo {author} {\bibfnamefont {R.~B.}\ \bibnamefont
  {Bird}}, \bibinfo {author} {\bibfnamefont {C.~F.}\ \bibnamefont {Curtiss}},
  \bibinfo {author} {\bibfnamefont {R.~C.}\ \bibnamefont {Armstrong}}, \ and\
  \bibinfo {author} {\bibfnamefont {O.}~\bibnamefont {Hassager}},\ }\href@noop
  {} {\emph {\bibinfo {title} {Dynamics of Polymeric Liquids}}},\ Vol.~\bibinfo
  {volume} {2}\ (\bibinfo  {publisher} {John Wiley and Sons},\ \bibinfo
  {address} {New York},\ \bibinfo {year} {1987})\BibitemShut {NoStop}%
\bibitem [{\citenamefont {Hockney}\ \emph {et~al.}(1973)\citenamefont
  {Hockney}, \citenamefont {Goel},\ and\ \citenamefont {Eastwood}}]{HGE}%
  \BibitemOpen
  \bibfield  {author} {\bibinfo {author} {\bibfnamefont {R.}~\bibnamefont
  {Hockney}}, \bibinfo {author} {\bibfnamefont {S.}~\bibnamefont {Goel}}, \
  and\ \bibinfo {author} {\bibfnamefont {J.}~\bibnamefont {Eastwood}},\ }\href
  {\doibase 10.1016/0009-2614(73)80315-X} {\bibfield  {journal} {\bibinfo
  {journal} {Chem. Phys. Lett.}\ }\textbf {\bibinfo {volume} {21}},\ \bibinfo
  {pages} {589 } (\bibinfo {year} {1973})}\BibitemShut {NoStop}%
\bibitem [{\citenamefont {Griebel}\ \emph {et~al.}(2007)\citenamefont
  {Griebel}, \citenamefont {Knapek},\ and\ \citenamefont {Zumbusch}}]{griebel}%
  \BibitemOpen
  \bibfield  {author} {\bibinfo {author} {\bibfnamefont {M.}~\bibnamefont
  {Griebel}}, \bibinfo {author} {\bibfnamefont {S.}~\bibnamefont {Knapek}}, \
  and\ \bibinfo {author} {\bibfnamefont {G.}~\bibnamefont {Zumbusch}},\
  }\href@noop {} {\emph {\bibinfo {title} {Numerical Simulation in Molecular
  Dynamics}}}\ (\bibinfo  {publisher} {Springer},\ \bibinfo {address} {Berlin,
  Heidelberg},\ \bibinfo {year} {2007})\BibitemShut {NoStop}%
\bibitem [{\citenamefont {Grest}\ \emph {et~al.}(1989)\citenamefont {Grest},
  \citenamefont {D{\"u}nweg},\ and\ \citenamefont {Kremer}}]{vectorMD}%
  \BibitemOpen
  \bibfield  {author} {\bibinfo {author} {\bibfnamefont {G.~S.}\ \bibnamefont
  {Grest}}, \bibinfo {author} {\bibfnamefont {B.}~\bibnamefont {D{\"u}nweg}}, \
  and\ \bibinfo {author} {\bibfnamefont {K.}~\bibnamefont {Kremer}},\ }\href
  {\doibase 10.1016/0010-4655(89)90125-2} {\bibfield  {journal} {\bibinfo
  {journal} {Comp. Phys. Comm.}\ }\textbf {\bibinfo {volume} {55}},\ \bibinfo
  {pages} {269 } (\bibinfo {year} {1989})}\BibitemShut {NoStop}%
\bibitem [{\citenamefont {Prabhakar}(2005)}]{prabhakarthesis}%
  \BibitemOpen
  \bibfield  {author} {\bibinfo {author} {\bibfnamefont {R.}~\bibnamefont
  {Prabhakar}},\ }\emph {\bibinfo {title} {Predicting the rheological
  properties of dilute polymer solutions using bead-spring models: {B}rownian
  dynamics simulations and closure approximations}},\ \href@noop {} {Ph.D.
  thesis},\ \bibinfo  {school} {Monash University} (\bibinfo {year}
  {2005})\BibitemShut {NoStop}%
\end{thebibliography}

%

\end{document}